\documentclass[11pt]{article}

\usepackage{url}
\usepackage[square,numbers]{natbib}

\usepackage{fullpage}

\usepackage{bm}
\usepackage{graphicx}
\usepackage{booktabs} %
\usepackage{hyperref}

\usepackage{amsmath}
\usepackage{amssymb}
\usepackage{mathtools}
\usepackage{amsthm}

\usepackage[capitalize,noabbrev]{cleveref}

\usepackage{thmtools} 
\usepackage{thm-restate}

\theoremstyle{plain}
\newtheorem{theorem}{Theorem}[section]
\newtheorem{proposition}[theorem]{Proposition}
\newtheorem{fact}[theorem]{Fact}
\newtheorem{lemma}[theorem]{Lemma}

\newtheorem{corollary}[theorem]{Corollary}
\theoremstyle{definition}
\newtheorem{definition}[theorem]{Definition}
\newtheorem{assumption}{Assumption}

\theoremstyle{remark}

\crefname{assumption}{Assumption}{Assumptions}
\crefname{finding}{Finding}{Findings}

\usepackage[inline]{enumitem}

\usepackage{amsmath,amsthm,amsfonts,amssymb,bm,bbm,mathtools,nicefrac}
\newcommand{\K}{\kappa}
\newcommand{\calM}{{{\mathcal{M}}}}
\newcommand{\calL}{{{\mathcal{L}}}}

\newcommand{\pos}{{\mathrm{pos}}}
\newcommand{\linf}{{{\lim_{m\rightarrow \infty}}}}

\usepackage{mathtools}

\usepackage{hyperref}
\usepackage{cleveref}
\allowdisplaybreaks

\usepackage{subcaption}

\usepackage{pgfplots}
\pgfplotsset{compat = 1.3}

\pgfplotsset{ylabel near ticks,every tick label/.append style={font=\Large},label style={font=\LARGE}
}

\newcommand{\appsymb}{$\bigstar$}
\newcommand{\normphi}{{{\mathrm{norm}\hbox{-}\phi}}}
\newcommand{\swap}{\mathrm{swap}}
\newcommand{\poso}{\mathrm{pos 1}}
\newcommand{\topo}{\mathrm{top 1}}
\renewcommand{\topo}{\mathrm{pos(c_1)} = 1}
\usepackage[textsize=tiny]{todonotes}

\usepackage{wrapfig}

\title{Properties of the Mallows Model Depending on the Number of
  Alternatives:\\ A Warning for an Experimentalist\footnote{This is an
    extended version of a paper presented at the 40th International
    Conference on Machine Learning (ICML-2023).}}

\author{
Niclas Boehmer\\
Technische Universit\"at Berlin\\
Berlin, Germany
\and
Piotr Faliszewski\\
AGH University\\
Krakow, Poland
\and
Sonja Kraiczy\\
University of Oxford\\
Oxford, Great Britain
}

\begin{document}

\maketitle

\begin{abstract}
The Mallows model is a popular distribution for ranked data. 
We empirically and theoretically analyze how the properties of rankings sampled from the Mallows model change when increasing the number of alternatives. 
We find that real-world data behaves differently than the Mallows model, yet is in line with its recent variant proposed by \citet{DBLP:conf/ijcai/BoehmerBFNS21}. 
As part of our study, we issue several warnings about
using the model.
\end{abstract}

\section{Introduction}

The Mallows model \cite{mallows1957non} is among the simplest and the most popular means of
generating and explaining %
ranking data. Indeed, it is even referred to as the normal
distribution over permutations.
The model has two main components, the central order (of the
available alternatives) and the dispersion parameter $\phi \in [0,1]$.
Depending on the value of $\phi$, it either generates random rankings
that are more concentrated around the central one or are more evenly
spread over the space of all permutations.
For example, the central order may rank the athletes participating
in some competition with respect to their expected performance, and
we can use the Mallows model to generate a realistic set of results in
a series of contests.  Using values of~$\phi$ close to $0$ means that
the differences in the strengths of the athletes are large and it is
unlikely that a weaker one would outperform a stronger one, whereas
using values of~$\phi$ close to $1$ means that their abilities are
similar and even the weakest participants may win some of the
contests.

This ability to generate realistic data is among the prime
applications of the Mallows model (another one is understanding
ranking data by estimating the Mallows parameters that are most likely
to lead to it). While real-life datasets typically involve fairly
small sets of alternatives and contain limited numbers of rankings,
synthetic models can provide arbitrarily large ranking profiles (i.e.,
collections of rankings over the same alternatives).  This flexibility
is useful, e.g., when evaluating various algorithms, such as
those
for aggregating search results~\cite{DBLP:conf/www/DworkKNS01},
analyzing elections~\cite{DBLP:journals/aamas/BetzlerBN14}, preference learning~\cite{DBLP:journals/jmlr/LuB14}, distribution testing \cite{busa2021identity}, 
clustering~\cite{DBLP:conf/icml/BusseOB07}, and for many other settings.
Yet, preparing synthetic datasets can be tricky.
To illustrate this, let us consider two common scenarios.

In the first one, we want to evaluate how the results of an algorithm
depend on the number of alternatives.
The most obvious approach here is to use the same, fixed dispersion
parameter for all the generated profiles and numbers of
alternatives. This approach is taken, e.g., in the works of
\citet{DBLP:conf/uai/MeilaPPB07,ali-mei:j:kemeny,DBLP:journals/jmlr/LuB14,bra-hof-str:c:no-show-paradox,aya-ben-lan-pet:c:stv-incomplete-votes,aya-ben:j:approximating-truncated-rules,cha-del-kim-kol-rel-sto:j:possible-necessary-winner,busa2021identity},
and others.
It is
so natural that it essentially never calls for any 
justification.  We conjecture that it is based on 
the
following, implicit assumption (needed to interpret the results).

\begin{assumption}\label{pass1}
  Using the Mallows model with a fixed dispersion parameter $\phi$ for
  different numbers of alternatives produces ``structurally similar''
  ranking profiles.
\end{assumption}

In the second scenario, we assume that there is no prior knowledge
regarding the structure of the ranking profiles that we want to
evaluate our algorithm on. 
In such a case, a very common approach is
to simply choose the dispersion parameter uniformly at random, either
from the $[0,1]$ interval, some subinterval of it, or from some
discrete set of equally-spaced values, such as $\{0.7,0.8, 0.9,1\}$.
This strategy is taken, e.g., by
\citet{DBLP:journals/jmlr/LuB14,busa2014preference,busa2021identity,bac-lev-lew-zic:c:misrepresentation,aya-ben-lan-pet:c:stv-incomplete-votes,ben-sko:c:truncated-votes}, and others.
As in the previous scenario, the approach is viewed to be so natural as to not need any justification. We
conjecture that the researchers make the following underlying assumption.

\begin{assumption}\label{pass2}
  Using the Mallows model with the dispersion parameter $\phi$ chosen
  uniformly at random
  (approximately) uniformly covers the space between identity
  profiles, where all rankings are equal, and profiles where each
  ranking is selected uniformly at random.
\end{assumption}

Yet, \citet{DBLP:conf/ijcai/BoehmerBFNS21} provided an interpretation
under which these two assumptions are false:
Regarding \Cref{pass1}, they found that as we fix the dispersion
parameter and increase the number alternatives, the Mallows model
generates rankings that, on average, become more and more similar
to the central one (where the similarity is measured in terms of the
swap distance, i.e., the number of pairs of alternatives ranked
differently by the two rankings, normalized by the total number of
pairs of alternatives).
Regarding \Cref{pass2}, their %
observation implies that if we choose the dispersion parameter
uniformly at random, then the more alternatives we consider, the more
the distribution is skewed toward choosing identity profiles.
They concluded by introducing a normalized variant of the dispersion
parameter, referred to as norm-$\phi$, which under their
interpretation satisfies the two assumptions and which they used in
their %
subsequent works
\cite{DBLP:journals/corr/abs-2205-07831,DBLP:conf/eaamo/BoehmerBFN22,DBLP:journals/corr/abs-2209-08856,DBLP:conf/aaai/BoehmerHN22,DBLP:journals/corr/abs-2208-04041,DBLP:conf/aaai/FaliszewskiSS22,DBLP:journals/corr/abs-2202-05115}.

If one accepts their interpretation (or, if their interpretation is
supported by real-life data) then quite a number of experiments based
on the Mallows model become questionable. Indeed, this goes even
beyond our two scenarios. For example, if one learned a realistic
value of $\phi$ using data with one number of alternatives, but used
it to produce test data with a different number of alternatives, then under the
interpretation of \citet{DBLP:conf/ijcai/BoehmerBFNS21} this generated
data is structurally different.

The purpose of this paper is to analyze and compare, theoretically and experimentally, how the properties of the Mallows model depend on the number of alternatives in case we use the classic or the normalized dispersion parameter.  %
We find that both variants 
have properties
that may be natural in some settings, but the data we consider points
toward the normalized variant.  Proofs of statements marked with
(\appsymb) can be found in the appendix.
The code for our experiments is available at \url{github.com/Project-PRAGMA/Normalized-Mallows-ICML-2023}.

\paragraph{Related Literature.}
The Mallows model has applications in many different fields, and we provide a brief overview of those relevant to the machine learning community. 
Motivated by the belief that user preferences can be understood as being sampled from the Mallows model or one of its variants \cite{ceberio2015review,chierichetti2018mallows,fligner1986distance}, the problem of fitting the parameters of the Mallows model
has been studied extensively, e.g.,  when user preferences are given as strict rankings  \cite{awasthi2014learning,liu2018efficiently,DBLP:journals/corr/abs-2205-07831}, incomplete rankings \cite{DBLP:conf/icml/ChengHH09,collas2021concentric,DBLP:conf/nips/LebanonM07,DBLP:conf/icml/BusseOB07}, or pairwise comparisons \cite{DBLP:journals/jmlr/LuB14,busa2014preference}. 
In addition, different algorithms for this task with applications in crowdsourcing \cite{busa2014preference}, recommendation systems \cite{DBLP:journals/jmlr/SunLK11,DBLP:conf/nips/LebanonM07}, and clustering \cite{DBLP:conf/icml/BusseOB07} have been implemented \cite{irurozki2016permallows,lee2013r}.
Additionally, the Mallows model has also proven useful in the context of Estimation of Distribution Algorithms for permutation-based problems \cite{ceberio2011introducing} applied in the field of evolutionary computation \cite{ceberio2015review}.
Motivated by the variety of applications, many properties of the Mallows model have already been studied, including the cycle structure \cite{gladkich2018cycle}, the longest increasing subsequence of a sampled ranking \cite{mueller2013length} and the thermodynamic limit \cite{starr2009thermodynamic}.

\section{Preliminaries} \label{sub:prelims}

A ranking profile $E = (C,V)$ consists of a set
$C = \{c_1, \ldots, c_m\}$ of alternatives and a collection
$V = (v_1, \ldots, v_n)$ of rankings (our notation largely follows
that used in the voting literature~\cite{DBLP:reference/choice/2016}, where a ranking profile would
typically be called an \emph{election}). Each ranking (sometimes also
called an order or a vote) is a strict, total order over $C$
that ranks the alternatives from the best to the
worst. %
We write $\calL(C)$ to denote the set of all rankings over alternative
set~$C$.  For a ranking $v\in \calL(C)$ and an alternative $c\in C$,
let $\pos(v,c)$ be the position of $c$ in $v$; the first alternative
has position $1$, the next one has position~$2$, and so on.
Given two rankings $u,v\in \calL(C)$, we write $\K(u,v)$ to denote
their swap distance, i.e., the number of pairs of distinct
alternatives $c, d \in C$ whose relative ranking is different in $u$
and $v$ (i.e., in one of the rankings $c$ is ranked higher than $d$,
and in the other it is the
opposite). %
The maximum swap distance of two rankings over $m$ alternatives is
$\binom{m}{2}$.  

Under the popular Plurality voting rule, the
Plurality score of an alternative in a ranking profile is the number
of rankings where the alternative appears in the first position.  The
alternative with the highest Plurality score is called a Plurality
winner (we will use the Plurality voting rule to get simple aggregate features of ranking profiles).

\paragraph{Mallows Model.}
The Mallows model $\calM_{\phi,m,v^*}$ is parameterized by a central
order $v^*\in \calL(C)$ over $m:=|C|$ alternatives, and a
dispersion parameter $\phi\in [0,1]$.
The probability of sampling a ranking $v\in \calL(C)$ under
$\calM_{\phi,m,v^*}$~is
$
  \textstyle \frac{1}{Z(\phi,m)} \phi^{\K(v^*,v)},
$
where $Z(\phi,m)$ is a normalization constant known to be
$(1+\phi)\cdot (1+\phi+\phi^2)\cdot \ldots \cdot(1+\phi+\ldots +\phi^{m-1})$.
Consequently, for $\phi=0$ only the central order $v^*$ is
sampled, whereas using $\phi=1$ leads to a uniform distribution over
rankings from $\calL(C)$, also known as \emph{Impartial Culture (IC)}.

In the following, we fix the central order $v^*$ to order the alternatives
lexicographically, i.e., to rank $c_1$ first, then $c_2$, and so on.
Hence, we will often write $\calM_{\phi,m}$ instead of
$\calM_{\phi,m,v^*}$.  The following fact gives the probability that
alternative $c_1$ appears in the $i$th position in a sampled ranking.
\begin{fact}[\citet{awasthi2014learning}]\label{le:c1} For all $i\in [1,m]$, it holds that
  $\textstyle \mathbb{P}_{v\sim
    \calM_{\phi,m}}[\pos(v,c_1)=i]=\frac{\phi^{i-1}}{\sum_{j=1}^{m}\phi^{j-1}}.$
\end{fact} %

\subsection{Measuring Properties of Mallows Model}
We are interested in measuring various properties of rankings sampled
from the Mallows model $\calM_{\phi,m}$, such as, e.g., the
probability that $c_1$ is ranked first.  For this, let $X_{\phi,m}$ be
a random variable capturing some property $\mathcal{X}$ we are interested in.
We define its normalized expected value to be:
\[
  g^{\mathcal{X}}_{m}(\phi)=\frac{\mathbb{E}[X_{\phi,m}]-\inf_{\phi'\in [0,1]} \mathbb{E}[X_{\phi',m}] }{\sup_{\phi'\in [0,1]}  \mathbb{E}[X_{\phi',m}]-\inf_{\phi'\in [0,1]} \mathbb{E}[X_{\phi',m}]}.
\]
For instance, for the probability that $c_1$ is ranked first,
\Cref{le:c1}, together with the observation that for $\phi=0$, $c_1$
is ranked on the first position with probability $1$ and for $\phi=1$
it is ranked first with probability $\nicefrac{1}{m}$, implies that
this function is:
\[
  \textstyle
g_m^{\topo}(\phi)=\frac{m}{m-1}\cdot
\left(\frac{1}{\sum_{j=1}^{m}\phi^{j-1}}-\frac{1}{m}\right).
\]
Let us assume that $g_m^{\mathcal{X}}$ is a bijection\footnote{For all properties $\mathcal{X}$ that we consider in our theoretical analysis, the expected value $\mathbb{E}[X_{\phi,m}]$  is strictly monotonic with respect to $\phi$. This is sufficient for $g_m^{\mathcal{X}}$
to be bijective.} and define
$\phi_m^{\mathcal{X}}:=(g^{\mathcal{X}}_m)^{-1}$. %
This function gives the dispersion parameter that leads to the
requested normalized expected value of $X_{\phi,m}$.
We say that we parameterize the Mallows model by property~$\mathcal{X}$ (or more precisely by $g_m^{\mathcal{X}}$, the normalized expected value of $X_{\phi,m}$) 
if instead of specifying the dispersion parameter~$\phi$ explicitly,
we specify the value $\ell \in [0,1]$ (of $g_m^{\mathcal{X}}$)
and use dispersion parameter~$\phi_m^{\mathcal{X}}(\ell)$ to sample from $\calM_{\phi_m^{\mathcal{X}}(\ell),m}$.

\subsection{Normalized Mallows Model} \label{sec:normphi}

Let us consider the random variable equal to the swap distance between
the sampled ranking and the central one (for $\phi=0$ its expected
value is $0$; for $\phi=1$ it is $m(m-1)/4$).  We denote its
normalized expected value as:
\begin{equation}\label{eq:norm-swap}
  g^{\swap}_m(\phi)=\frac{4\cdot \mathbb{E}_{v\sim
      \calM_{\phi,m,v^*}}[\K(v,v^*)]}{m(m-1)}.
\end{equation}
Using the terminology from the previous section, the normalized
Mallows model of \citet{DBLP:conf/ijcai/BoehmerBFNS21} is simply the
Mallows model parameterized by the normalized expected swap distance
(between the sampled ranking and the central order).
When we use this expected value as a free variable, then---following
\citet{DBLP:conf/ijcai/BoehmerBFNS21}---we denote it by $\normphi$ and
refer to it as the normalized dispersion parameter\footnote{In
  \Cref{strict} in Appendix \ref{sec:swapap}, we show its
  well-definedness, i.e., that each $\normphi$ gives rise to a unique
  $\phi$, which was missing in the work of
  \citet{DBLP:conf/ijcai/BoehmerBFNS21}.}
(note that at the end of the preceding section, $\normphi$ would take
the role of $\ell$).
Accordingly, using the normalized Mallows model, one specifies a value of $\normphi\in [0,1]$, which is then internally converted to a corresponding value $\phi$  of the dispersion parameter such that $g_m^{\swap}(\phi)=\normphi$ (see \Cref{sec:norm-Mallows}). 
Rankings are then sampled from $\calM_{\phi,m}$ (using standard algorithms such as the Repeated Insertion Model; see the start of \Cref{sec:Deleting} for a description).

To get an exact formula for $g^{\swap}_m(\phi)$, it suffices to
replace $\mathbb{E}_{v\sim \calM_{\phi,m,v^*}}[\K(v,v^*)]$ in
\eqref{eq:norm-swap} with the following result.

\begin{fact}[\citet{diaconis2000analysis}, Property~4] \label{f1}
  Given dispersion parameter $\phi\in [0,1)$, the expected swap
  distance between the central order and a sampled one is:
  \[ \textstyle
    \mathbb{E}_{v\sim \calM_{\phi,m,v^*}}[\K(v,v^*)]=\frac{m\phi}{1-\phi}-\sum_{i=1}^{m}i\frac{\phi^i}{1-\phi^{i}}.
  \]
\end{fact}

\begin{figure}
  \centering
    \resizebox{0.35\textwidth}{!}{\input{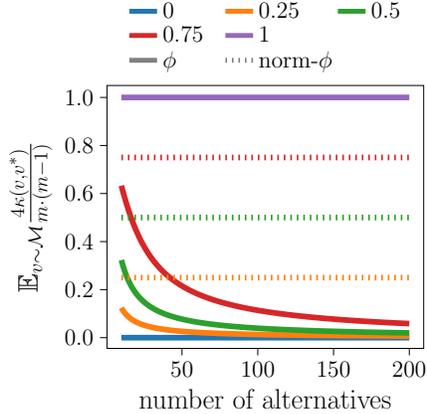}}
    \caption{Expected normalized swap distance of a sampled ranking
      from the central one (solid lines for the classic model and
      dashed ones for the normalized variant). }\label{fig:Exp}
\end{figure}
Using \Cref{f1} we are also able to explain the experimental
observation of \citet{DBLP:conf/ijcai/BoehmerBFNS21} that for a fixed dispersion parameter sampled
rankings become more and more similar to the central one as we
increase the number of alternatives. Indeed, for a fixed dispersion parameter
$\phi\in [0,1)$, $\mathbb{E}_{v\sim \calM_{\phi,m,v^*}}[\K(v,v^*)]$
grows at most linearly in~$m$, but the denominator
in~\eqref{eq:norm-swap} is $m(m-1)$. Hence, the normalized expected
swap distance goes to $0$ as $m$ grows.
\begin{corollary}\label{Cor:zero}
    For fixed $\phi<1,
   \lim_{m\to \infty} g^{\swap}_m(\phi)=0$.%
\end{corollary}
To visualize the speed of convergence and the difference between using
the classic and normalized models, in \Cref{fig:Exp} we
show how the expected normalized swap distance changes for fixed
values of $\phi$ and $\normphi$. %

\section{Mallows Versus Normalized
  Mallows} \label{sec:fixvsnorm}

In this section we provide our main comparison of the classic and
normalized variants of the Mallows model. In particular, we consider a
number of properties---such as, e.g., the expected position of $c_1$
in a sampled ranking---and we evaluate if for a fixed (normalized) dispersion parameter value the property is almost 
independent of the number of alternatives, or if for all values of the (normalized) dispersion parameter it converges to
the same constant as the number of alternatives grows. 

We view our properties as measuring various structural
properties of the sampled rankings (or profiles) and if a given
property does not depend strongly on the number of alternatives (for a
given variant of the Mallows model), then we say that from the
perspective of this property, \Cref{pass1} holds (for this variant).
To better understand the two Mallows models, we first build
some intuitions regarding their behavior in \Cref{sec:Deleting}, and
then, in \Cref{sec:norm-Mallows}, we analyze how $\phi$ relates to
$\normphi$.  Afterward (\Cref{sec:asy_cover}), we introduce our framework for analyzing the
properties and then perform the analysis (\Cref{sec:c1,sec:PC,sec:Winners}).

\subsection{Intuitions for the Two Models} \label{sec:Deleting}

\begin{figure}
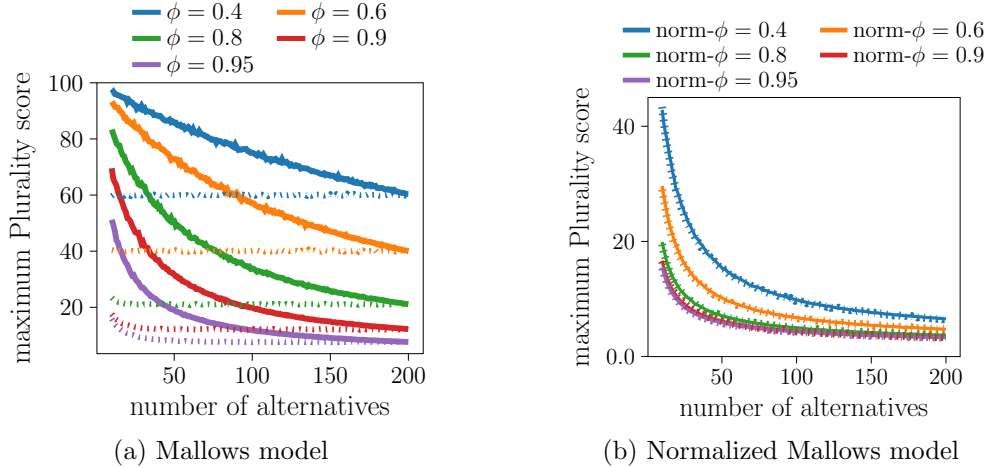

  \centering
    \begin{subfigure}{.35\textwidth}
  \centering
   \resizebox{\textwidth}{!}{\input{plots/deletions/maxPlur.tex}}
  \caption{Mallows model}
  \label{fig:del1}
\end{subfigure}
\quad\quad\quad\quad
\begin{subfigure}{.35\textwidth}
  \centering
  \resizebox{\textwidth}{!}{\input{plots/deletions_normphi/maxPlur.tex}}
  \caption{Normalized Mallows model}
  \label{fig:del2}
\end{subfigure}
\caption{Average Plurality score of Plurality winner in $100$ rankings
  with a varying number of alternatives. We compare sampling profiles
  with a varying number of $m$ (dashed) with sampling profiles for
  $m=200$ alternatives and subsequently deleting some alternatives
  uniformly at random (solid).} \label{fig:del}
\end{figure}

It will be helpful to consider the classic Mallows model through the
lenses of the Repeated Insertion
Model (\citet{diaconis2000analysis}, Property 3; see \citet{DBLP:journals/jmlr/LuB14}, Section 2.2.3 for a more accessible description). The idea is
that to sample a ranking from $\calM_{\phi,m}$, we can first sample a
ranking from $\calM_{\phi,m-1}$, which regards one alternative fewer,
and then insert the missing alternative $c_m$ at some position~$j$.
Specifically, the probability of inserting $c_m$ at position
$j \in [m]$ is $\frac{\phi^{m-j}}{\sum_{i=0}^{m-1}\phi^i}$.  After doing so, we can
also insert alternatives $c_{m+1}$, $c_{m+2}$, and so on, if we so
choose and want to obtain a ranking over more alternatives.  Hence, sampling a ranking from the Mallows model is an
iterative procedure that inserts the alternatives into the sampled ranking one by one, following the central order.
In other words, we can imagine the sampling process as first sampling a ranking over
many more than $m$ alternatives, and then restrict it to 
the top $m$ from the central order.

On the other hand, for the normalized Mallows model %
it is not possible to iteratively expand rankings like this, as for a fixed value of the normalized dispersion parameter $\normphi$ the corresponding value of $\phi$ increases with increasing $m$.\footnote{Note however that in case the number of alternatives is known upfront, the Repeated Insertion Model can still be used to sample rankings from the normalized Mallows model (after converting $\normphi$ to the respective value of $\phi$).}  
However, we can use the following intuition: 
Instead of sampling a ranking of $m$ alternatives, we can sample a ranking over a large
number of alternatives, but then restrict it to \emph{randomly
  selected} $m$ ones.  Since there is no theory that would back this
point of view, we perform the following experiment. For each number
$m$ of alternatives ranging between $20$ and $200$, we first sample a
profile of $100$ rankings according to the (normalized) Mallows model
for $200$ alternatives, and then we delete a random subset of them so
that only $m$ remain. Then we compute the Plurality score of the
Plurality winner.\footnote{We view the Plurality score of the
  Plurality winner as an example of a feature of a ranking profile. In
  \Cref{app:Deleting}, we show similar results for other properties. }
Finally, we repeat this experiment for the case where we sample a profile
of $100$ rankings over exactly $m$ alternatives (without
any deleting). We show the results in \Cref{fig:del}. We see that both
experiments give nearly identical results for the normalized Mallows
model (supporting our intuitive view of it), but vary greatly for the
classic one.

Looking at it from a yet different perspective, when increasing the number of alternatives in the classic Mallows model, we add an alternative at the end of the central order, leaving the relation between all the other alternatives intact. 
In contrast, for the normalized Mallows model, we add an alternative and then move all alternatives ``closer together,'' in the sense of increasing the probability that each pair of them may be swapped during sampling (for a fixed value of $\normphi$, the respective value of $\phi$ increases when increasing~$m$). 
It is sometimes also helpful to view this as inserting an alternative of random quality, in contrast to inserting an alternative worse than all already present ones as in the classic Mallows model.

\subsection{Relation Between $\boldsymbol\phi$ and $\boldsymbol\normphi$} \label{sec:norm-Mallows} To analyze rankings %
sampled from the normalized Mallows model, we want to better
understand the relation between~$\phi$ and~$\normphi$.  While we
already presented a formula for $g^{\mathrm{swap}}_m(\phi)$ in
\Cref{sub:prelims} (allowing to move from $\phi$ to $\normphi$),
a closed formula for the other direction seems elusive.
Nonetheless, 
we provide an asymptotic result.

\begin{restatable}[\appsymb]{theorem}{swapphi}
\label{thm:swapphi}
Fix $\ell\in [0,1]$. Then,
$$\lim_{m\rightarrow \infty} (1-\phi_m^{\mathrm{swap}}(\ell))\times m=h^{\swap}(\ell),$$

where $h^{\swap}(\ell)$ is the unique solution $L$ to $\int_{0}^1 \gamma(s,L) ds=\frac{c}{4}$ with $\gamma:(0,1]\times \mathbb{R}\rightarrow \mathbb{R}, (s,x)\mapsto \frac{1}{x}-s\frac{1}{e^{sx}-1}$. Furthermore,
 $h^{\swap}$ is a bijective strictly decreasing function from $(0,1]$ to $[0,\infty)$, 
 
\end{restatable}
Note that for fixed $\ell\in [0,1]$, the above implies that
$\phi_m^{\mathrm{swap}}(\ell)$ behaves asymptotically like
$1-\frac{h^{\swap}(\ell)}{m}$. %
However, to sample rankings from the normalized
model, in the absence of a closed form expression for
$\phi^{\mathrm{swap}}_m$, we need to find a different way to convert
values of $\normphi$ to values of $\phi$.  As
$g^{\mathrm{swap}}_m(\phi)$ is strictly monotonic (see \Cref{strict} in the
appendix), given some value of $\normphi$, we can simply perform a
binary search on $\phi\in [0,1]$ until
$g^{\mathrm{swap}}_m(\phi)=\normphi$.

\subsection{On Covering a Property Asymptotically} \label{sec:asy_cover}
In the remainder of this section, we want to analyze 
how various properties of sampled rankings behave as we change
the number of alternatives, while keeping $\phi$ or $\normphi$ fixed.
Specifically,
in our theoretical analysis, we focus on the case where the number of
alternatives goes to infinity.

\begin{definition}\label{def:normalizes}
  We say that parameterizing by property $\mathcal{Y}$
  \emph{asymptotically covers} property $\mathcal{X}$ if there is a
  bijective and strictly monotonic function $f: [0,1]\rightarrow [0,1]$
  such that for all $\ell\in [0,1]$:
  $\linf g_{m}^{\mathcal{X}}(\phi_{m}^{\mathcal{Y}}(\ell)) =f(\ell).$

  Alternatively, we say that parameterizing by property $\mathcal{Y}$
  \emph{asymptotically cannot distinguish} property $\mathcal{X}$ if
  for all $\ell\in (0,1)$:
  $\linf g_{m}^{\mathcal{X}}(\phi_{m}^{\mathcal{Y}}(\ell)) =L,$ for some constant
  $L$.
\end{definition}
Intuitively speaking, if parameterizing by property $\mathcal{Y}$
asymptotically cannot distinguish property $\mathcal{X}$, then for all
$\ell\in (0,1)$, when keeping a fixed value $\ell$ of $\mathcal{Y}$ (i.e.,
we select some $\ell\in (0,1)$ and sample from
$\calM_{m,\phi_m^{\mathcal{Y}}(\ell)}$), property $\mathcal{X}$ converges
to the same value as $m$ goes to infinity.  Practically speaking, this
means that as the number of alternatives increases, sampled rankings
become more and more similar with respect to the range of potential
values of property~$\mathcal{X}$.  In contrast, if $\mathcal{Y}$
asymptotically covers $\mathcal{X}$, then there is a ``well-behaved''
mapping $f$ between all possible expected values of $\mathcal{X}$ and
all possible expected values of $\mathcal{Y}$ such that if we
parameterize the Mallows model by a fixed value $\ell$ of $\mathcal{Y}$
and increase the number of alternatives, the value of property
$\mathcal{X}$ converges to $f(\ell)$. 
Consider the expected swap distance from the central order as an illustrative toy property (cf. \Cref{fig:Exp}). Parameterizing by the dispersion parameter asymptotically cannot distinguish this property, whereas (by definition) parameterizing by the normalized dispersion parameter asymptotically covers it.

Notably, for all ``well-behaved'' properties our two notions are
symmetric.
\begin{restatable}[\appsymb]{proposition}{symmetry}
Let $\mathcal{X}$ and $\mathcal{Y}$ be properties such that $g^{\mathcal{X}}_m(\phi)$ and $g^{\mathcal{Y}}_m(\phi)$ are strictly monotonic and continuous.\footnote{These conditions hold for all properties that we consider.}
  If parameterizing by property $\mathcal{Y}$ asymptotically covers
  (cannot distinguish) property $\mathcal{X}$, then parameterizing by
  property $\mathcal{X}$ asymptotically covers (cannot distinguish)
  property $\mathcal{Y}$.
\end{restatable}
Thus, two parameterizations are ``similar'' if they asymptotically
cover each other, while they are different if they asymptotically
cannot distinguish each other.

\begin{figure}[t]
\centering
\begin{subfigure}{.35\textwidth}
  \centering
  \resizebox{\textwidth}{!}{\input{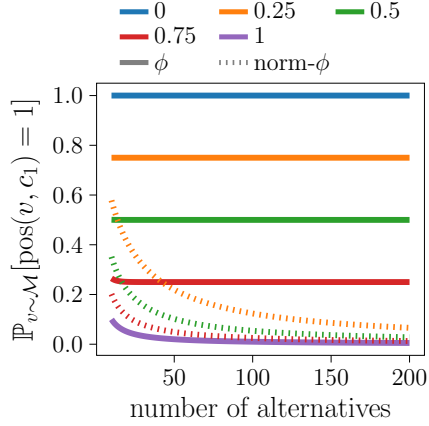}}
  \caption{Probability that $c_1$ is ranked first in a sampled vote.}
  \label{fig:elecsub1}
\end{subfigure}\quad\quad\quad
\begin{subfigure}{.35\textwidth}
  \centering
  \resizebox{\textwidth}{!}{\input{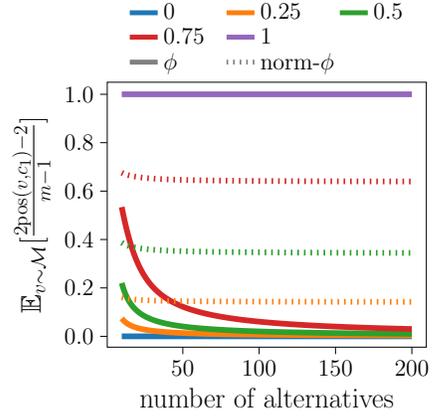}}
  \caption{Expected normalized position of $c_1$ in a sampled vote.}
  \label{fig:elecsub2}
\end{subfigure}\hfill

\begin{subfigure}{.35\textwidth}
  \centering
   \resizebox{\textwidth}{!}{\input{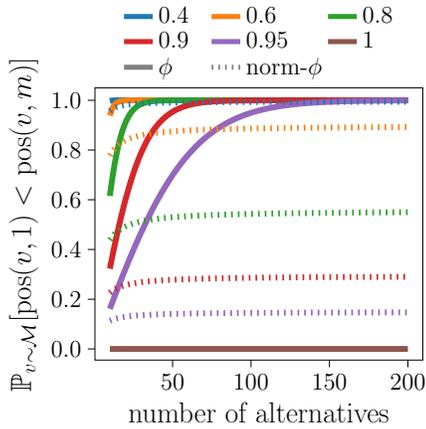}}
  \caption{Probability that $c_1$ is ranked before~$c_m$.}
  \label{fig:sub5}
\end{subfigure}
\quad\quad\quad
\begin{subfigure}{.35\textwidth}
  \centering
  \resizebox{\textwidth}{!}{\input{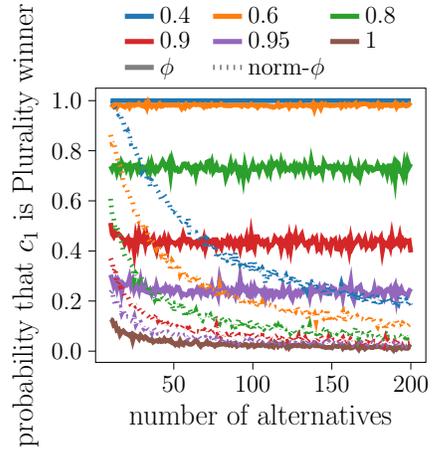}}
  \caption{Fraction of profiles with $c_1$ as Plurality winner.}
  \label{fig:elecsub3}
\end{subfigure}
\label{fig:elections}
\caption{Influence of the number of alternatives $m$ on different properties of rankings (ranking profiles) sampled from the Mallows model for fixed values of the classical dispersion parameter $\phi$ (solid) and the normalized dispersion parameter $\normphi$ (dashed). For $\phi=\normphi=0$ and $\phi=\normphi=1$ the respective lines overlap. }
\end{figure}

While asymptotic coverage of some property $\mathcal{X}$ does not
directly imply that for the respective variant of the Mallows model
property $\mathcal{X}$ stays constant for a fixed parameter value when
changing the number of alternatives, it is clearly a necessary
condition.  Moreover, we observe empirically for all considered
variants and properties that asymptotic coverage of some property
corresponds to the property staying approximately constant in
practice.

In the following, we will focus on parameterizing by the dispersion
parameter (thereby recovering the classic Mallows model) and by the
expected swap distance (thereby recovering the normalized Mallows
model).
Accordingly, we will say that the normalized/classic Mallows model asymptotically covers or cannot distinguish some property if this holds for the respective parameterization.
In fact, it turns out that
if the classic model asymptotically covers some property, then
the normalized one cannot asymptotically distinguish it, and the other
way round.

\begin{restatable}[\appsymb]{theorem}{implications} \label{th:impli}
Let $\mathcal{X}$ be a property such that $g^{\mathcal{X}}_m(\phi)$ is strictly monotonic:
    \setlist{nolistsep}
    \begin{enumerate}[noitemsep]
        \item If the normalized Mallows model asymptotically covers property $\mathcal{X}$, the classic Mallows model asymptotically cannot distinguish~$\mathcal{X}$.\label{i1}
        \item If the classic Mallows model asymptotically covers property $\mathcal{X}$, then the normalized Mallows model asymptotically cannot distinguish $\mathcal{X}$. \label{i2} 
    \end{enumerate}
\end{restatable}

Hence, the classic and the normalized models indeed are very different
from each other, and, in particular, exhibit a fundamentally different behavior with respect to all properties considered in the paper.

\subsection{Position of the Central Order's Top-Choice} \label{sec:c1}
In this and the following sections we use our asymptotic
framework to analyze basic properties of sampled rankings and ranking
profiles. We first look at the position of $c_1$. Both of our models fix it, but in two different ways.

\paragraph{Probability That $\boldsymbol{c_1}$ Is Ranked First.} \label{sec:expPlurScore}
Consider the probability that $c_1$ is ranked first in a sampled
ranking, i.e., $\mathbb{P}_{v\in \calM}[\pos(v,c_1)=1]$. 
By \Cref{le:c1},
we know that it is
$\frac{1}{\sum_{i=0}^{m-1}\phi^i}$.  Since
$\sum_{i=0}^{m-1}\phi^i$ is a geometric series, we have that $\linf \sum_{i=0}^{m-1}\phi^i=\frac{1}{1-\phi}$. 
From this, it is easy to conclude
the following.
\begin{theorem}
  The classic Mallows model asymptotically covers the probability
  that $c_1$ is ranked first, with $f(\ell)=1-\ell$.
\end{theorem}
In fact, as witnessed by \Cref{fig:elecsub1}, for a fixed value of $\phi$ the probability that $c_1$ is ranked first quickly converges to $1-\phi$.

\paragraph{Expected Position of $\boldsymbol{c_1}$.}

Next, let us consider the normalized  position of $c_1$, i.e.,
$g^{\poso}_m(\phi)=\frac{2\mathbb{E}_{v\sim
    \calM_{\phi,m}}[\pos(v,c_1)]-2}{m-1}$, which describes its average
performance.  We start by deriving a closed formula for this property.
\begin{restatable}[\appsymb]{proposition}{exppos}
\label{prop:exppos}The expected position of $c_1$ in a sampled ranking is $\mathbb{E}_{v\sim \calM_{\phi,m}}[\pos(v,c_1)]=\frac{1}{1-\phi}-m\frac{\phi^m}{1-\phi^m}.$
\end{restatable}

Given that we discussed above that fixing the dispersion parameter leads to a roughly constant probability of $c_1$ being ranked first, it is intuitive that this parameterization asymptotically cannot distinguish the expected normalized position of $c_1$.
In contrast, parameterizing by the swap distance can do so.

\begin{restatable}[\appsymb]{theorem}{expposnorm}
\label{prop:expposnorm} The normalized Mallows model asymptotically covers the expected position of $c_1$, with $f(\ell)=t(h^{swap}(\ell))$, where $t(x)=2\cdot (\frac{1}{x}-\frac{1}{e^{x}-1}).$
\end{restatable}
Again, as seen in \Cref{fig:elecsub2}, convergence is reached fairly quickly.
The behavior of $c_1$ in both  parameterizations is in line with
our discussion from \Cref{sec:Deleting}: In the classic model,
increasing the number of alternatives means adding them at the end of
the central order.  Thus, these new alternatives have an
exponentially shrinking probability of being ranked first, leading to
a convergence of the probability of~$c_1$ being ranked first.
For the normalized Mallows model, the intuition is that increasing the
number of alternatives means that there are more and more alternatives
similar to~$c_1$ (who then can end up ahead of $c_1$ with a non-negligible probability).

\subsection{Pairwise Comparisons}\label{sec:PC}
Next, we turn to the probability that some alternative $c_i$ is ranked
before some other alternative $c_j$.
\paragraph{Comparing Two Fixed Alternatives.}
\citet{mallows1957non} showed that the probability that $c_i$ is
ranked before $c_j$ depends only on the difference between $i$ and $j$
and is independent of the number of alternatives. Using this, we
calculate the probability that $c_i$ appears
above~$c_j$. %

\begin{restatable}[\appsymb]{proposition}{comparingTwo}\label{winprob}
    Consider $1\leq i,j\leq m$ with $k:=j-i+1$. The probability that $c_i$ is ranked before $c_j$ is 
   $$\frac{1}{1-\phi^{k}}\left(1-\frac{(1-\phi)(k-1)\phi^{k-1}}{1-\phi^{k-1}}\right).$$
\end{restatable}
Notably, since for each $i>j\in \mathbb{N}$ $g_m^{i\text{ beats }j}(\phi)=2\cdot\mathbb{P}_{v\sim \calM_{\phi,m}}[\pos(v,c_i)<\pos(v,c_j)]$ does not depend on $m$, we can say that the classic Mallows model ``covers'' the probability that $c_i$ is ranked before $c_j$ (for each $m$) and, so, clearly also  covers it asymptotically.

\paragraph{Comparing $\boldsymbol{c_1}$ and  $\boldsymbol{c_m}$.}
If instead of comparing the relative ranking
of %
two alternatives with fixed indices, we compare the probability that
$c_1$ is ranked before $c_m$, the picture changes (see \Cref{fig:sub5}).
\begin{restatable}[\appsymb]{theorem}{promm}
\label{th:pro1m}
The normalized Mallows model asymptotically covers the probability that alternative  $c_1$ is ranked before alternative  $c_m$ in a sampled ranking, with 
$f(\ell)=t(h^{\swap}(\ell))$, where $t(x)=2\cdot\frac{1}{1-e^{-x}}(1-\frac{x}{e^{x}-1})-1.$
\end{restatable}

Our discussions from \Cref{sec:Deleting} explain the behavior of the classic model: As we add alternatives at the
end of the central order, these added alternatives are viewed as
more and more inferior to $c_1$ (so the probability of placing them before $c_1$ in a sampled ranking goes to $0$). 
In contrast, under the normalized model, the
quality difference between $c_1$ and $c_m$ stays roughly constant.

\subsection{Winners in Ranking Profiles} \label{sec:Winners}
More complex properties may concern an entire profile, not just an individual ranking. %
For example, the results from \Cref{sec:expPlurScore} show that the classic Mallows model asymptotically covers the expected Plurality score of $c_1$ (because it is equal to the probability that $c_1$ is ranked first, times the number of sampled rankings).
For a constant number of votes, we present empirical evidence in \Cref{fig:elecsub3} that the classic Mallows model also asymptotically covers the probability that $c_1$ is the Plurality winner. 
In \Cref{app:winners}, we show that 
the same seems to hold
for more involved voting rules such as Borda and Condorcet.

\section{Examining Real-World Evidence}
\label{sub:evidence}
Next, we evaluate how a notion analogous to the expected swap distance
from the central order depends on the number of alternatives in
real-world data (in addition, in \Cref{app:real-worldData}, we examine
the properties of alternative $c_1$ as discussed in the previous
section\footnote{In real-world
  profiles it is not straightforward who the strongest alternative $c_1$
  is.  This is why we compare the behavior of rankings
  sampled from the Mallows model and real-world rankings considering
  the Plurality winner (as a proxy for $c_1$).}).
We analyze the following three datasets of
\citet{DBLP:journals/corr/abs-2204-03589}:\footnote{ To the best of
  our knowledge, the work of \citet{DBLP:journals/corr/abs-2204-03589}
  is the only one containing complete profiles with a wide spectrum of
  numbers of alternatives; we selected the three datasets from their
  paper with the highest variation of alternative
  numbers.} %
\begin{description} %
\item[American Football.] In each week in a season of American College
  Football, different media outlets publish their %
  rankings of the teams by estimated strength.  Each American football profile regards one
  week with rankings %
  published in this week.
  However %
  there are two divisions in college football (FBS and FCS) with some
  outlets ranking all teams from the FBS division and some outlets
  ranking all teams from both division.  Accordingly, we created two
  different profiles for each week, one for rankings over FBS teams and one for rankings over FBS
  and FCS teams.  Thus the profiles from this dataset have different
  sizes (also the numbers of teams in each division vary slightly
  between years).

\item[Spotify.] Spotify profiles are based on the daily top $200$
  songs on Spotify in different countries. Each profile in this
  dataset regards a single month in a single country, and contains a
  single ranking for each day of this month.
  As a Spotify profile may be incomplete, we delete all alternatives
  that do not appear in all rankings.

\item[Tour de France.] Tour de France is a multistage bike race held
  annually in France. Each profile corresponds to one edition, and each
  ranking orders the cyclists in a single stage, where we delete all cyclists that do not finish all stages. 
   Tour
  de France profiles are of different sizes because the number
  of participants differs in each year (from $67$ to $252$) and
  because the fraction of cyclists finishing all races varies
  (which sometimes reduces the number of cyclists even below $20$).
\end{description}
Notably, all three datasets may be regarded as ground-truth-based,
implying that on a conceptual level the Mallows model is a suitable
or, at least, a natural choice to capture them.\footnote{In
  American Football profiles, in the considered week, there is some
  underlying ranking of the teams by strength and each outlet tries to
  estimate the ground truth using their own (principled) method.  In a
  Spotify profile, one could argue that there is an underlying
  popularity ranking of songs in each month, yet the streams of
  songs are subjected to fluctuation on different days leading to
  small perturbations of the underlying ranking.  In Tour de France
  profiles, it is usually assumed that there is some true ordering of
  cyclists by strength, and that we sample perturbed versions of that
  ordering by doing races on different days. }
We will see that ranking profiles from the above datasets behave
differently than the classic Mallows model, but mostly in line with the
normalized variant.

\begin{figure*}[!tb]
    \centering
    \begin{minipage}[b]{1\textwidth}
        \centering
\begin{subfigure}[b]{.32\textwidth}
  \centering
   \resizebox{\textwidth}{!}{
\begin{tikzpicture}

\definecolor{color0}{rgb}{0.12156862745098,0.466666666666667,0.705882352941177}

\begin{axis}[
tick align=outside,
tick pos=left,
x grid style={white!69.0196078431373!black},
xlabel={number of alternatives},
xmin=107.893346774194, xmax=264.106653225806,
xtick style={color=black},
y grid style={white!69.0196078431373!black},
ylabel={positionwise distance from ID},
ymin=0, ymax=1,
ytick style={color=black},
ytick={0,0.2,0.4,0.6,0.8,1},
yticklabels={0.0,0.2,0.4,0.6,0.8,1.0}
]
\addplot [draw=color0, fill=color0, mark=*, only marks,mark size=1.5pt]
table{%
x  y
236 0.174134123350391
115 0.222600322645695
236 0.182689812207722
115 0.241515426497278
236 0.169016967411796
115 0.267071234119782
236 0.203195978094982
115 0.243868641781527
236 0.219802401236056
115 0.20630534015289
236 0.190749944911001
115 0.175177363471374
236 0.172680286765932
115 0.157453619681387
236 0.155730160774716
115 0.13740373767597
237 0.121116561031192
115 0.140297738755089
237 0.116568362140469
115 0.134163779354342
237 0.116392419380514
115 0.120505331215971
237 0.115852023760651
115 0.124479555887691
237 0.115695324740066
115 0.127117362371446
240 0.176161912533204
117 0.21500584453536
240 0.202803000052084
117 0.247530683810637
240 0.212526837850194
117 0.256812536528346
240 0.210211982846924
117 0.224378588372813
240 0.246366456682608
117 0.20645821157218
240 0.184990190801924
117 0.185384165804972
240 0.177821374213673
117 0.169947055385567
240 0.161570513376968
117 0.15228667445938
240 0.161730238372194
117 0.150361911612642
240 0.148414034965885
117 0.147326125073057
240 0.149186617823226
117 0.13775641117019
240 0.150873452664109
117 0.133658205253763
240 0.150152606816091
117 0.135562459626565
240 0.147965068838001
117 0.127026361930542
240 0.146530321706974
117 0.124327109415854
240 0.225301530532524
117 0.221917007597896
240 0.227087275820761
117 0.211225160724722
240 0.223954780960979
117 0.234771000478189
240 0.233319001657504
117 0.209543451872499
240 0.21855587770621
117 0.178701886949988
240 0.200396471529651
117 0.17156633547633
240 0.17384416396118
117 0.149693161893629
240 0.164815361377802
117 0.136084692630572
240 0.165727877220091
117 0.128677186830314
240 0.16061621440187
117 0.124317078748761
240 0.154605307576713
117 0.118653199883109
240 0.152051250889772
117 0.11833357685564
240 0.147764911042051
117 0.114567802573981
240 0.143108883338748
117 0.113863237872589
240 0.165877879824303
117 0.211815702318332
240 0.189308842167399
117 0.229076563413209
240 0.185387833923399
117 0.226994447691409
240 0.203262457433537
117 0.256023118384311
240 0.227516204836072
117 0.259755878253833
240 0.194922365550328
117 0.219856173608111
237 0.151318770959848
117 0.204935556307484
237 0.149925224327019
117 0.194385593220339
237 0.194498920819958
117 0.209422225933038
237 0.225513934387314
117 0.231899473991818
237 0.208792794548685
117 0.223582700175336
239 0.185536522301228
119 0.200484261501211
239 0.19014939309057
119 0.23215799031477
239 0.211840569561158
119 0.25
239 0.227704831932773
119 0.236440677966102
241 0.167787377911345
119 0.199118644067797
241 0.199298996458087
119 0.255524168236033
241 0.222623966942149
119 0.260825587752871
241 0.236452184179457
119 0.243267419962335
241 0.2356222654351
119 0.240084745762712
241 0.190946656649136
120 0.204458642961317
241 0.194249311294766
120 0.224152470905718
241 0.205075757575757
120 0.220937961961644
241 0.227341597796143
120 0.20726439336065
241 0.227578512396694
120 0.213934127369956
245 0.190332200453152
120 0.195357769410315
245 0.204347672695627
120 0.264218348496423
245 0.187748430039749
120 0.235679946909893
245 0.185750699720112
120 0.2517150110733
245 0.211756206608266
120 0.254103294210246
245 0.198776739304278
120 0.217576837897693
245 0.218719654995145
120 0.217127147239631
245 0.209823213571714
120 0.229534169039517
245 0.215980274556844
120 0.250490368137427
245 0.212898174063708
120 0.239710524115195
244 0.166214831611657
120 0.186923900366644
244 0.197184429327286
120 0.202160503570324
244 0.20119425547997
120 0.203568058239338
244 0.200534139581759
120 0.22477751431151
244 0.240366687985736
120 0.242144008538605
244 0.219666302735086
120 0.208733400847696
244 0.204525069286974
120 0.197708877462482
244 0.180315627934676
120 0.176058974285138
246 0.197119819587754
120 0.211694808555254
246 0.218821779724036
120 0.216937108087703
246 0.211735544420197
120 0.223550877901875
246 0.224987901464808
120 0.250127019054686
246 0.234972805260594
120 0.251666782415446
246 0.157170105530368
120 0.16107179160169
246 0.155007023052136
120 0.162825093306377
246 0.162459048208305
124 0.151050064184852
246 0.198227955149671
124 0.216975609756098
246 0.212173580278052
124 0.222817073170732
246 0.223288440882426
124 0.217827050997783
246 0.219627903969192
124 0.218483849703362
246 0.226086094356771
124 0.237429480381761
246 0.207634024175378
124 0.232
246 0.205013275689967
124 0.213078590785908
246 0.191328220651529
124 0.195843902439024
246 0.178017667197138
124 0.185864773078111
246 0.172590338723062
124 0.178331707317073
246 0.170583649759851
124 0.171175166297117
246 0.169126590963793
124 0.169117516629712
246 0.16251192416606
124 0.158967582587218
246 0.16215598719685
124 0.158113821138211
246 0.162606422532834
124 0.155370731707317
252 0.195481578298138
125 0.204070660522273
252 0.204184054296648
125 0.209400921658986
252 0.209078986144636
125 0.205085125448029
252 0.218047897771348
125 0.220046082949309
252 0.234100301223114
125 0.228086597542243
252 0.135971528904146
128 0.144742420099686
252 0.189228405407169
128 0.216339236665184
252 0.210784192422855
128 0.227430873466398
252 0.212579792221699
128 0.23754937609543
252 0.205469238358475
128 0.225215162058231
252 0.242445238807615
128 0.259053709520657
252 0.215542913273729
128 0.220194608793388
252 0.188485583358266
128 0.211416014858608
252 0.170019829468989
128 0.187486360318372
252 0.158824779931657
128 0.176636350688722
252 0.151894398689826
128 0.169609961545505
252 0.146737949388218
128 0.163083208679253
252 0.143652133968704
128 0.160659032111147
252 0.142922381619766
128 0.155672929890106
252 0.141315390131367
128 0.158926223017049
252 0.135721147032424
128 0.157366164324597
252 0.137145376708862
128 0.150716588745512
253 0.138547648649182
128 0.131344575585557
253 0.163432695913011
128 0.210291155466032
253 0.19699100112486
128 0.220964239935124
253 0.205331241489551
128 0.209821980926174
253 0.209700037495313
128 0.212153713343971
253 0.212042460209715
128 0.220133847902623
253 0.202882905765812
128 0.196495974387298
253 0.185773653293338
128 0.178689531113672
253 0.169135930377124
128 0.171401998862749
253 0.158822252481598
128 0.160029298663249
253 0.151115978923687
128 0.153437292186702
253 0.147021096047205
128 0.151042319220501
253 0.14755392418053
128 0.148895235875642
253 0.144495096007736
128 0.141816798839322
253 0.140008413582448
128 0.142250717603217
253 0.140959514207065
128 0.140280749081067
253 0.141236457284945
128 0.155388930116906
253 0.168950920608608
128 0.187950924739059
253 0.199498500187477
128 0.225599707013368
253 0.192114378559823
128 0.215354107395679
253 0.216344484717188
128 0.220157309973895
253 0.213728218183254
128 0.218588955450336
253 0.197521362461271
128 0.202532241609334
253 0.177627186845547
128 0.187981609149893
253 0.170209973753281
128 0.182031679179637
253 0.15716703175261
128 0.176441607589749
253 0.150123126501079
128 0.170216616219826
253 0.147598326524974
128 0.164200475220788
253 0.144788644840448
128 0.165088811572972
253 0.142457521757149
128 0.163466485293989
253 0.141373734533183
128 0.162431083057739
254 0.133640755379886
130 0.136128617077934
254 0.171669379214136
130 0.225129388630187
254 0.182786492881272
130 0.248560777053587
254 0.200175669740887
130 0.23294991950146
254 0.199740924037599
130 0.231116418938184
254 0.219253196930946
130 0.230161548020593
254 0.190632336065881
130 0.201013730663516
254 0.17734316232887
130 0.182192494688558
254 0.164987400793027
130 0.170124711521392
254 0.152711057058883
130 0.165351365599901
254 0.148475906020663
130 0.164727811948497
254 0.145108770172709
130 0.162587065715688
254 0.136533929884313
130 0.159360141252561
254 0.133532253481103
130 0.158211517744018
254 0.135164749998509
130 0.15347426734337
254 0.132248314345501
130 0.148691451689761
255 0.133789270323473
130 0.151956274981306
255 0.163478100393701
130 0.226611042073496
255 0.200275702397996
130 0.250714692360903
255 0.209684116633858
130 0.237014636654689
255 0.22197342519685
130 0.240734165729728
255 0.219276301399825
130 0.227760828011015
255 0.204462967519685
130 0.208885948065781
255 0.185885211614173
130 0.199992433347982
255 0.17160257311586
130 0.188751161746431
255 0.166084610517435
130 0.173467576272476
255 0.15425776856018
130 0.173414711596403
255 0.15160812417979
130 0.166765291831864
255 0.139539427975915
130 0.167126692265977
255 0.140812183633296
130 0.159320442889139
255 0.134529173843504
130 0.158357496498807
255 0.139934194775484
130 0.154796760030842
256 0.134282983049022
130 0.147261460948661
256 0.195990467599262
130 0.220714666125637
256 0.219811832166821
130 0.245729371274116
256 0.217812097191939
130 0.220611166821002
256 0.216965852488855
130 0.224182587044295
256 0.205458915083543
130 0.219190071794533
256 0.203205973117813
130 0.204284900071944
256 0.189901579308766
130 0.19430553469254
256 0.183746667244035
130 0.184493165276052
256 0.174499631240304
130 0.178660383347051
256 0.160161236998042
130 0.174546817365919
256 0.154570592565557
130 0.164933503024181
256 0.147362560546355
130 0.16336645539557
256 0.144841179013759
130 0.156798575681093
256 0.139881633587287
130 0.157580921948044
256 0.14008276988364
130 0.154688640353473
256 0.139846392512907
130 0.157271515668595
257 0.155265867248062
130 0.262950470442038
};
\end{axis}

\end{tikzpicture}}
  \caption{American Football}
  \label{fig:ammm}
\end{subfigure}\hfill
\begin{subfigure}[b]{.32\textwidth}
  \centering
  \resizebox{\textwidth}{!}{\input{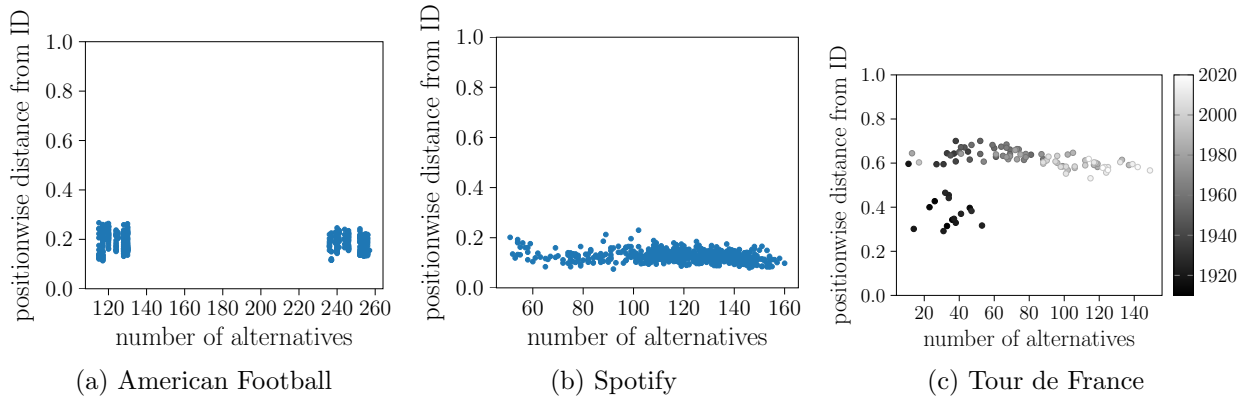}}
  \caption{Spotify}
  \label{fig:spooo}
\end{subfigure}\hfill
\begin{subfigure}[b]{.34\textwidth}
  \centering
  \resizebox{\textwidth}{!}{
\begin{tikzpicture}
\pgfkeys{/pgf/number format/.cd,
int detect,
1000 sep={\,},
min exponent for 1000 sep=4}
\begin{axis}[
colorbar,
colorbar style={ylabel={}},
colormap/blackwhite,
legend style={fill opacity=0.8, draw opacity=1, text opacity=1, draw=white!80!black},
point meta max=2020,
point meta min=1910,
tick align=outside,
tick pos=left,
x grid style={white!69.0196078431373!black},
xlabel={number of alternatives},
xmin=4.09334677419355, xmax=155.906653225806,
xtick style={color=black},
y grid style={white!69.0196078431373!black},
ylabel={positionwise distance from ID},
ymin=0, ymax=1,
ytick style={color=black},
ytick={0,0.2,0.4,0.6,0.8,1},
yticklabels={0.0,0.2,0.4,0.6,0.8,1.0}
]
\addplot [colormap/blackwhite, forget plot, only marks, scatter, scatter src=explicit]
table [x=x, y=y, meta=colordata]{%
x  y  colordata
33 0.314705882352941 1910.0
26 0.427259259259259 1911.0
36 0.342857142857143 1912.0
23 0.4 1913.0
46 0.396784869976359 1914.0
11 0.596666666666667 1919.0
14 0.301538461538462 1920.0
38 0.329313929313929 1921.0
31 0.291666666666667 1922.0
37 0.347368421052632 1923.0
53 0.316809116809117 1924.0
41 0.370039682539683 1925.0
34 0.455003819709702 1926.0
32 0.465542521994135 1927.0
34 0.441558441558441 1928.0
47 0.382905138339921 1929.0
27 0.59483994266603 1933.0
31 0.595052083333333 1934.0
37 0.64327485380117 1935.0
33 0.6445842760181 1936.0
38 0.7003267003267 1937.0
45 0.652173913043478 1938.0
38 0.607708300015992 1939.0
43 0.670995670995671 1947.0
35 0.634920634920635 1948.0
46 0.615873015873016 1949.0
41 0.672077922077922 1950.0
52 0.700517943026267 1951.0
40 0.650297740435599 1952.0
61 0.628445747800587 1953.0
52 0.641416944136145 1954.0
60 0.667533910934856 1955.0
81 0.638056733828208 1956.0
47 0.682301512287335 1957.0
69 0.661239495798319 1958.0
59 0.68205329153605 1959.0
69 0.640679908326967 1960.0
64 0.674392274392274 1961.0
88 0.617691154422789 1962.0
67 0.68406660184735 1963.0
67 0.669674688057041 1964.0
77 0.61441647597254 1965.0
65 0.6328125 1966.0
74 0.662537224538416 1967.0
54 0.606689536878216 1968.0
73 0.663429054054054 1969.0
89 0.610072951739619 1970.0
78 0.641259675932213 1971.0
78 0.626978964898613 1972.0
68 0.617624918883842 1973.0
87 0.640338266384778 1974.0
69 0.631460084033613 1975.0
67 0.62722816399287 1976.0
41 0.64271978021978 1977.0
61 0.639516129032258 1978.0
73 0.617065929565929 1979.0
72 0.659270692649045 1980.0
99 0.586168831168831 1981.0
13 0.644557823129252 1982.0
75 0.639504165819955 1983.0
99 0.585640074211503 1984.0
119 0.582569337442219 1985.0
102 0.638915347845464 1986.0
105 0.646748004354136 1987.0
117 0.591696585121482 1988.0
119 0.597055084745763 1989.0
133 0.61472184531886 1990.0
137 0.591097308488613 1991.0
102 0.568509083918101 1992.0
112 0.612117473784917 1993.0
96 0.601139446554531 1994.0
17 0.603070175438596 1995.0
106 0.578167715684404 1996.0
118 0.601163542340013 1997.0
88 0.597481596280511 1998.0
112 0.589787132265008 1999.0
96 0.632816060770483 2000.0
90 0.612161345455254 2001.0
106 0.584526737404259 2002.0
95 0.607502799552072 2003.0
101 0.583931888544892 2004.0
98 0.583536394876601 2005.0
88 0.603448275862069 2006.0
91 0.629202898550725 2007.0
101 0.552296918767507 2008.0
117 0.60333138515488 2009.0
132 0.604235780290421 2010.0
124 0.556956097560976 2011.0
125 0.581778033794163 2012.0
139 0.595326086956522 2013.0
115 0.531133869155648 2014.0
123 0.581973823373876 2015.0
149 0.566756756756757 2016.0
142 0.582141262992326 2017.0
125 0.568490783410138 2018.0
122 0.578653497278774 2019.0
114 0.618963337547408 2020.0
};
\end{axis}

\end{tikzpicture}}
  \caption{Tour de France}
  \label{fig:tdddfff}
\end{subfigure}
\caption{Plots showing the normalized positionwise distance of a profile from ID depending on the number of alternatives in the profile. Each point corresponds to one profile. For Tour de France, the color of a point corresponds to the year of the respective edition.}
\label{fig:disFromID-real}
    \end{minipage}
\end{figure*}

\subsection{Positionwise Distance From Identity} \label{sub:posDisID}%

While in a real-life dataset there is no central order, in principle
we could estimate it.  For example, the normalized Kemeny score of a
profile $(C,V)$ is defined as
$\min_{v^*\in \calL(C)}\sum_{v\in V} \K(v,v^*)/(|V|{|C| \choose 2})$
and the ranking $v^*$ that achieves this minimum is the Kemeny order
(technically, it does not need to be unique, but it is not crucial for
our discussion). The Kemeny order is a maximum-likelihood estimator
for the central order of the Mallows model producing the given
profile~\cite{MM09}, and the normalized Kemeny score is the normalized
swap distance of the profile from this ranking.
Unfortunately, computing the Kemeny score of a profile is
NP-hard~\cite{bartholdi1989voting}.  Thus, we turn to a
polynomial-time approximation known as the ``positionwise distance
from ID,''\footnote{Here, each ranking profile is modeled by a
  frequency matrix, where we have one column for each alternative and
  one row for each position, and an entry contains the fraction of
  rankings that rank the alternative on this position.  The distance
  between two frequency matrices is then defined as the minimum earth
  mover's distance between columns over all possible column mappings.
  The positionwise distance from ID of a profile is the distance of
  the profile's frequency matrix from the identity matrix.} proposed
by \citet{DBLP:conf/ijcai/BoehmerBFNS21}.  Its values range between
$0$ and $1$ and have a similar interpretation as the normalized Kemeny
score (in particular, value $0$ means that all rankings in the profile
are identical and value~$1$ means that the rankings are maximally
diverse). Consequently, in this section we use it as a moral
equivalent of the normalized expected swap distance from the central
ranking that we would have used, had we been looking at Mallows data.

\begin{figure}
  \centering
  \resizebox{0.35\textwidth}{!}{\input{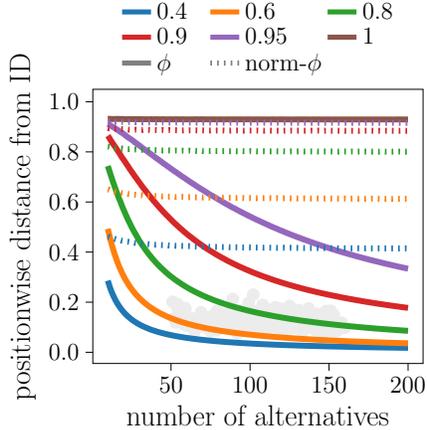}}
  \caption{Average positionwise distance from ID of profiles with
    $n=100$ rankings (see \Cref{sub:posDisID}). Lightgrey points are
    from \Cref{fig:spooo}.}\label{fig:disID2}
\end{figure}

Before examining the positionwise distance from ID of real-world
ranking profiles, in \Cref{fig:disID2}, we show its behavior on
profiles sampled from the classic and normalized Mallows models; note the intuitive connection to the expected swap distance
(\Cref{fig:Exp}).
\Cref{fig:disFromID-real} shows the normalized positionwise distance
from ID of profiles from American Football, Spotify, and Tour de
France.  We see here that in all three datasets, the
positionwise distance from ID stays constant when varying the number
of alternatives.  We interpret this as evidence that
real-world profiles that come from the same source
do not behave like those from  the classic Mallows model with a fixed dispersion
parameter (for which the normalized expected swap distance from the
central order goes to $0$ as the number of alternatives increases), but rather like those from the normalized variant (where it stays fixed).

Moreover, this evidence highlights a practical problem with using the
classic Mallows model: The Kemeny score and the positionwise distance
from ID can both be
used to
estimate the dispersion parameter of the Mallows model (see, e.g., the works
of~\citet{MM09} and~\citet{DBLP:journals/corr/abs-2205-07831}).
This estimated dispersion
parameter might then, for instance, be employed to generate more similar
data of varying sizes to conduct further experiments.  If the
classic dispersion parameter is used, then this approach is
problematic: For instance, assume that we have access to Spotify
profiles with between $50$ and $60$ alternatives. This leads to an
estimated dispersion parameter around $0.6$; however, if we use this
parameter to generate profiles with a higher number of alternatives,
then we get data different from the true Spotify profiles with larger
alternative numbers, as for $m=160$ the estimated dispersion parameter
of Spotify profiles is around $0.8$ (see lightgrey points in
\Cref{fig:disID2}). Note further that reporting a value of $\phi=0.8$
for Spotify profiles with $m=160$ and a value of $\phi=0.6$ for
Spotify profiles with $m=50$, one might also suggest that the nature
of Spotify elections changes when varying the number of alternatives,
while this behavior is rather a feature of the Mallows model.

\section{A Final Takeaway: Be Cautious}
We analyzed how the Mallows model behaves as we vary the
number of alternatives and argued that for the classic variant
this behavior is unnatural.  Instead, we generally suggest to use the normalized Mallows model
which keeps certain structural properties of sampled rankings constant,
in line with real-world data.
While the normalized Mallows model seems to be advantageous in many contexts, we want to remark that in works with specific applications in mind one should base one's decisions on available data (in particular, on its dependence on the number of alternatives). We looked at three data sets that are better captured by the normalized Mallows model. 
However, we could also imagine applications where the classic Mallows model is more suitable, for instance, in recommendation settings (e.g., lists of the top-$x$ smartphones published on some websites). Here one could imagine that the classic Mallows model would be a better fit, as we believe that when we increase the number of alternatives (smartphones) in such lists, the newly added alternatives should be worse than the already ranked ones.

Independent of whether one agrees with our interpretation of the
results, we made several observations that should be taken into
account when conducting experiments with data generated from the
Mallows
model. %
\emph{First}, one should be extremely careful when conducting
experiments with a fixed dispersion parameter and varying $m$, as it
is unclear whether observed trends are because of the increased number
of alternatives or the changed structure of the profile.  Consider as an
example the setup used by \citet{busa2014preference}: They use data
generated from the Mallows model to compare their algorithm to a
baseline one.  Considering $\phi\in \{0.1,0.3,0.5,0.7\}$, they observe
for $m=10$ that their algorithm is better than the baseline algorithm
for small values of $\phi$. Then, they repeat their experiments with
$\phi\in \{0.1,0.3,0.5,0.7\}$ and $m=20$ and conclude that here ``the
advantage of [our algorithm] is even more pronounced'' (as their algorithm now outperforms the baseline one also for larger values of the dispersion parameter). Considering our
findings, it is not clear whether one can really conclude that their
algorithm has a stronger competitive advantage when increasing the
number of alternatives, or if for $m=20$ they looked at data which is simply more favorable for their algorithm. 
To illustrate this point, consider the following idealized example: 
There are two algorithms A and B, where A is better than B on data with a  ``low'' level of disagreement, say, on profiles where the Kemeny score is less than a third of the maximum possible. Then, as an example, for $10$ alternatives algorithm A is better than B for $\phi<0.4$ and for $20$ alternatives A is better than B for $\phi<0.6$. Using this observation, one might conclude that algorithm A scales better in the number of alternatives than algorithm B; however, the reason for this observed trend is that for $m=20$ the ``low range'' of the level of disagreement extends to the case where $\phi$ is below $0.6$ (and not only below $0.4$ as for $10$ alternatives).

\emph{Second}, statements about how an
algorithm behaves for a certain dispersion parameter or parameter
ranges might no longer be true when varying the number $m$ of
alternatives. 
Accordingly, one should clarify that it is unclear whether such statements generalize for different alternative numbers, even if one only conducts experiments with a fixed number of alternatives. 
\emph{Third}, one should be careful how to select the
values of the dispersion parameter used to generate data for
experiments to ensure a meaningful coverage of the space of ranking
profiles.  For instance, by picking values of $\phi$ uniformly at
random one might risk to mostly produce profiles where sampled
rankings are very similar and only barely cover the ranges in which
real-world data appears.  \emph{Fourth}, the above described problems
get intensified when considering generalizations of the Mallows model. For instance,
consider a mixture of two Mallows models with uniformly at random
sampled central orders.  Then increasing the number of alternatives
for a fixed value of $\phi$, the nature of the sampled profiles
change, as the two clusters around the central orders become more
and more clearly separated from each~other.

One should also be cautious when using the Mallows model to describe
or learn user preferences when faced with a varying number of alternatives. 
For instance, imagine a questionnaire given by a company
which asks first, how respondents would rank $5$ drafts of
conventional television ads and, second, how respondents would rank
$25$ drafts for more provocative ads. Later, the extensive theory of
learning Mallows parameters is used to estimate that the preferences
over traditional ads are best captured by a dispersion parameter of
$\phi=0.5$, whereas preference over the more provocative ones are best
captured by $\phi=0.9$. This might lead to the belief that opinions
concerning the provocative ads are more varied (and as the company
fears a public relations issue, they decide to stick with the most
popular traditional ad instead).  However, this interpretation is
questionable, as we argued that values of $\phi$ for
different numbers of alternatives are in certain ways incomparable.

\section*{Acknowledgments}
Niclas Boehmer is supported by the DFG project ComSoc-MPMS (NI 369/22).  This
project has received funding from the European Research Council (ERC)
under the European Union’s Horizon 2020 research and innovation
programme (grant agreement No 101002854).

\begin{center}
  \includegraphics[width=3cm]{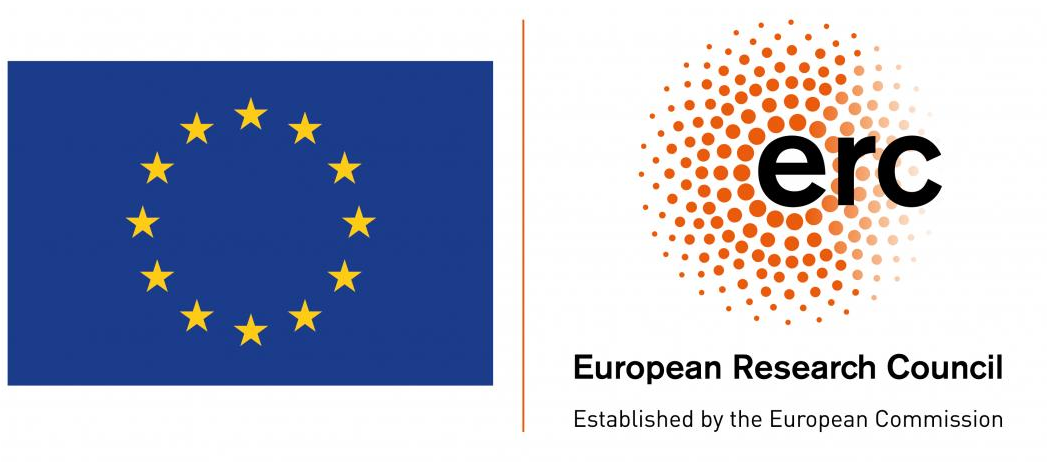}
\end{center}

\bibliography{bib}
\bibliographystyle{plainnat}

\newpage
\appendix
\onecolumn

\section{Additional Material for \Cref{sec:normphi}}\label{sec:swapap}
The following Lemma implies that $g^{swap}_m$ is strictly increasing and continuous, bijectively mapping $[0,1]$ to $[0,1]$.

\begin{lemma}\label{strict}
$\mathbb{E}_{v\sim \calM_{\phi,m,v^*}}[\K(v,v^*)]$ is strictly increasing and continuous in $\phi \in [0,1]$ and bijectively maps $[0,1]$ to $[0,\frac{1}{2}\cdot\binom{m}{2}]$.
\end{lemma}
\begin{proof}
Note that $\mathbb{E}_{v\sim \calM_{0,m,v^*}}[\K(v,v^*)]=0$ and $\mathbb{E}_{v\sim \calM_{1,m,v^*}}[\K(v,v^*)]=\frac{1}{2}\cdot\binom{m}{2}$. It remains to show that $\mathbb{E}_{v\sim \calM_{\phi,m,v^*}}[\K(v,v^*)]$ is continuous and strictly increasing on $[0,1]$.

As an immediate consequence of Property 3 in ~\cite{diaconis2000analysis} the expected swap distance can be decomposed as $\mathbb{E}_{v\sim \calM_{\phi,m,v^*}}[\K(v,v^*)]=\sum_{i=n}
^m \mathbb{E}[X_{\phi_,n}]$,  where $X_{\phi,n}$ represents the number alternatives stronger (i,e, ranked earlier in the central order) than the $n$th alternative are ranked below the $n$th alternative in a sample vote $v$.
The random variable $X_{\phi,n}$ has truncated geometric distribution  $G_{\phi,n}$, i.e., $\mathbb{P}(X_{\phi,n}=i)=\frac{\phi^{i-1}}{\sum_{j=1}^{n}\phi^{j-1}}$ for all $i\in[n]$, implying that $\mathbb{E}[X_{\phi,m}]=\frac{\sum_{j=1}^m j \phi^{j-1}}{\sum_{j=1}^{m}\phi^{j-1}}$.
To conclude continuity of $\mathbb{E}_{v\sim \calM_{\phi,m,v^*}}[\K(v,v^*)]$, simply observe that it is a sum of continuous function, since $\mathbb{E}[X_{\phi,m}]$ is the ratio of continuous functions with the denominator being non-zero.

We will show that $\mathbb{E}_{v\sim \calM_{\phi,m,v^*}}[\K(v,v^*)]$ is a sum of functions strictly increasing in $\phi$, so also strictly increasing itself. 
Let $0\leq \phi_1<\phi_2 \leq 1$.
We can express \begin{align*}\mathbb{E}[X_{\phi_2,m}]=\frac{\sum_{j=1}^m j \phi_1^{j-1}+\sum_{j=1}^m j (\phi_2^{j-1}-\phi_1^{j-1})}{\sum_{j=1}^m \phi_1^{j-1}+\sum_{j=1}^m (\phi_2^{j-1}-\phi_1^{j-1})}. \end{align*}
So since $\sum_{j=1}^m j (\phi_2^{j-1}-\phi_i^{j-1})>\sum_{j=1}^m (\phi_2^{j-1}-\phi_1^{j-1})$,  it follows that $\mathbb{E}[X_{\phi_1,m}]<\mathbb{E}[X_{\phi_2,m}]$.
So as desired, $\mathbb{E}_{v\sim \calM_{\phi,m,v^*}}[\K(v,v^*)]$ is striclty increasing and maps [0,1] bijectively to $[0,\frac{1}{2}\cdot\binom{m}{2}]$.

\end{proof}

\section{Additional Material for \Cref{sec:fixvsnorm}}
\subsection{ Missing proofs from \Cref{sec:norm-Mallows}}
\swapphi*
To prove Theorem \ref{thm:swapphi} we will need \Cref{lem:unique} %
and \Cref{hswap}. The proof of \Cref{lem:unique} in turn uses \Cref{technical}, \Cref{limalg} and \Cref{loglimit}, which we will prove first. Note that \Cref{loglimit} will also be used in proofs in later subsections.
 \begin{proposition} \label{technical}Let $\gamma:(0,1]\times \mathbb{R}\rightarrow \mathbb{R}, (s,x)\mapsto \frac{1}{x}-\frac{s}{e^{sx}-1}$. The following hold
 \begin{enumerate}
 \item For fixed $s\in(0,1]$, $\gamma(s,x)$ is strictly decreasing in $x$.
 \item $f$ is uniformly continuous%
     \item $\int_0^1 \gamma(s,0)ds=\int_0^1 \frac{1}{2}sds=\frac{1}{4}$
 \end{enumerate}
  \end{proposition}
 \begin{proof}
 We can reexpress $\gamma(s,x)$ as \begin{align*}\gamma(s,x)=\frac{1}{x}-\frac{s}{\sum_{i=0}^{\infty}\frac{(sx)^i}{i!}-1}=\frac{\sum_{i=0}^{\infty}\frac{(sx)^i}{i!}-sx-1}{x\cdot (\sum_{i=0}^{\infty}\frac{(sx)^i}{i!}-1)}=
\frac{\sum_{i=2}^{\infty}\frac{(sx)^{i-1}}{i!}}{x\cdot(\sum_{i=1}^{\infty}\frac{(sx)^{i-1}}{i!})}=\frac{\sum_{i=2}^{\infty}\frac{x^{i-2}s^{i-1}}{i!}}{\sum_{i=1}^{\infty}\frac{(sx)^{i-1}}{i!}}\end{align*}
 so $\gamma(s,x)$ is continuous and differentiable as it is the ratio of continuous and differentiable functions, and the function in the denominator always takes values $\geq 1$.
 For fixed $s\in (0,1]$, we compute the partial derivative with respect to $x$:
 \begin{align*}\frac{d}{dx}\gamma(s,x)=-\frac{1}{x^2}+\frac{s^2}{(e^{sx}-1)^2}=-\frac{1}{x^2}+\frac{1}{(x+\frac{sx^2}{2}+\frac{s^2x^3}{3!}+\ldots)^2}<-\frac{1}{x^2}+\frac{1}{x^2}=0,\end{align*}
 since $s>0$ so that $\gamma$ is strictly decreasing in $x$, proving \textbf{point 1}.
 Since $\lim_{x\rightarrow \infty}\gamma(s,x)=0$ and $\lim_{x\rightarrow -\infty}\gamma(s,x)=s$ and $f$ is strictly decreasing in $x$ it is bounded in $[0,s]\subset[0,1]$. From continuity and boundedness we conclude that $\gamma(s,x)$ is uniformly continuous, proving \textbf{point $2$}.
From our reformulation of $\gamma$, we see that $\lim_{x\rightarrow 0}f(x,s)=\frac{s}{2}$ so its integral along the unit interval is $\frac{1}{4}$, proving \textbf{point $3$}. %
 \end{proof}
  \begin{proposition}\label{limalg} Let $x,w \in \mathbb{R}$ and $s\in [0,1]$. Then
    \begin{align*}\frac{1}{x}-\frac{s}{e^{s(x+w)}-1}=\frac{ws}{x(w+x)\times\sum_{i=1}^{\infty} \frac{(s(x+w))^{i-1}}{i!}}+\frac{w+x}{x} \times f(s,x+w).\end{align*}
\end{proposition} 
    \begin{proof}
        \begin{align}&\frac{1}{x}-\frac{s}{e^{s(x+w)}-1}=\frac{e^{s(x+w)}-sx-1}{x\times(e^{s(x+w)}-1)}
        \\&=\frac{\sum_{i=0}^{\infty}\frac{(s(x+w))^i}{i!}-sx-1}{x\times \sum_{i=0}^{\infty}\frac{(s(x+w))^i}{i!}-1 } =\frac{ws+\sum_{i=2}^{\infty}\frac{(s(x+w))^i}{i!}}{x\times \sum_{i=1}^{\infty}\frac{(s(x+w))^i}{i!}}\label{alg3}\\
    &=\frac{w}{x(x+w)\times\sum_{i=1}^{\infty} \frac{(s(x+w))^{i-1}}{i!}}+\frac{w+x}{w+x}\frac{\sum_{i=2}^{\infty} \frac{(s(x+w))^{i-1}}{i!}}{x\times \sum_{i=1}^{\infty} \frac{(s(x+w))^{i-1}}{i!}},\\
 \end{align} 
    where Equation \ref{alg3} follows using the Maclaurin series for Eulers function and
    \begin{align}&\frac{w+x}{w+x}\frac{\sum_{i=2}^{\infty} \frac{(s(x+w))^{i-1}}{i!}}{x\times \sum_{i=1}^{\infty} \frac{(s(x+w))^{i-1}}{i!}}
    \\&=\frac{w+x}{x}\frac{\sum_{i=2}^{\infty} \frac{(s(x+w))^{i-1}}{i!}}{(w+x)\times \sum_{i=1}^{\infty} \frac{(s(x+w))^{i-1}}{i!}}
    \\&=\frac{w+x}{x}\frac{e^{s(x+w)}-s(x+w)-1}{(w+x)\times e^{s(x+w)}-1}\\&=\frac{w+x}{x}(\frac{1}{(w+x)}-\frac{s}{e^{s(w+x)}-1})  =\frac{w+x}{x}f(s,x+w),\end{align}
     so $\frac{1}{x}-\frac{s}{e^{s(x+w)}-1}=\frac{ws}{x(w+x)\times\sum_{i=1}^{\infty} \frac{(s(x+w))^{i-1}}{i!}}+\frac{w+x}{x} \times f(s,x+w)$, as desired.
    \end{proof}

\begin{lemma}\label{loglimit}
    If $\linf (1-\phi_m)\times m =L$, then $\linf \log(\phi_m^{-m})=L.$
\end{lemma}
\begin{proof}
    Since $\linf (1-\phi_m)\times m =L$, $\linf (1-\phi_m) =0$ and so
     \begin{align}&\linf \log(\phi_m^{-m}) =\linf - m \log(\phi_m)=\lim_{m\rightarrow \infty}m\times \sum_{i=1}^\infty \frac{(1-\phi_m)^i}{i}\nonumber\\&=\lim_{m\rightarrow \infty}m\times (1-\phi_m) +\sum_{i=2}^\infty m(1-\phi_m)\frac{(1-\phi_m)^{i-1}}{i}\nonumber=L,\end{align}
     using the Maclaurin series of the logarithm.
\end{proof}
\begin{lemma}\label{lem:unique}
    For $c\in [0,1]$. Suppose $\phi_m$ is a sequence in $[0,1]$, satisfying 
$$\lim_{m\rightarrow \infty} (1-\phi_m)\times m =L\text{ and }\linf g^{swap}_m(\phi_m)=c.$$
Then $L$
is the unique solution to $4\int_{0}^1\gamma(s,L) ds=c$ with $f:(0,1]\times \mathbb{R}\rightarrow \mathbb{R}, (s,x)\mapsto \frac{1}{x}-s\frac{1}{e^{sx}-1}$.
\end{lemma}
\begin{proof}

Let $x_{m}=(1-\phi_m)\times m$ and let $w_{m}=\sum_{i=2}^\infty m(1-\phi_m)\frac{(1-\phi_m)^{i-1}}{i}$. %
By \Cref{loglimit}, we know that $\linf \log(\phi_m^{-m})=\linf x_m+w_m=L$.

    Let $s(i)=\frac{i}{m}\in[0,1]$ for $1\leq i \leq m$. 
    Then 
    \begin{align}\label{reform}
      \phi_m^{-i}=e^{-i\log(\phi_m)}=e^{\frac{i}{m}\times (-m\log(\phi_m ))}=e^{s(i) \times (x_{m}+w_{m})}.
    \end{align}

    \begin{align}&\frac{c}{4}=\frac{1}{4}\cdot\linf g^{swap}_m(\phi_m)= \linf \frac{\mathbb{E}_{v\sim \calM_{\phi_m,m,v^*}}[\K(v,v^*)]}{m^2}
    \\&=\linf\frac{\phi_m }{m(1-\phi_m)}-\frac{1}{m^2}\sum_{i=1}^{m}i\frac{1}{\phi_m^{-i}-1})
    \\&=\linf(\frac{\phi_m-1}{m(1-\phi_m)} +\frac{1 }{m(1-\phi_m)}-\frac{1}{m^2}\sum_{i=1}^{m}i\frac{1}{\phi_m^{-i}-1})
    \\&=0+\linf( \frac{1 }{x_{m}}-\frac{1}{m^2}\sum_{i=1}^{m}i\frac{1}{\phi_m^{-i}-1})
    \\&=\lim_{m\rightarrow\infty}(\frac{1}{m}\sum_{i=1}^{m} \frac{1}{x_{m}}-\frac{i}{m}\frac{1}{\phi_{c}(m)^{-i}-1})\label{phiver}
    \\&=\lim_{m\rightarrow\infty}\frac{1}{m}(\sum_{i=1}^{m} \frac{1}{x_{m}}-s(i)\frac{1}{e^{s(i)(x_{m}+w_{m})}-1})\label{seqver}
\end{align}
   Equation \ref{phiver} equals Equation \ref{seqver} using Equation \ref{reform}. It remains to show the first equality in the following, as the second follows by definition of the Riemann integral.
   \begin{align}    
    \lim_{m\rightarrow\infty}\frac{1}{m}(\sum_{i=1}^{m} \frac{1}{x_{m}}-s(i)\frac{1}{e^{s(i)(x_{m}+w_{m})}-1})=\lim_{m\rightarrow \infty} \frac{1}{m}\sum_{i=1}^{m} \gamma(s(i),L)=\int_{0}^1 \gamma(s,L) ds\label{integral}.
   \end{align}
   By \Cref{limalg}, $$\frac{1}{x_{m}}-\frac{s}{e^{s(x_{m}+w_{m})}}=\frac{w_ms}{x_m(w_m+x_m)\times\sum_{i=1}^{\infty} \frac{(s(x_m+w_m))^{i-1}}{i!}}+\frac{w_m+x_m}{x_m} \times \gamma(s,x_m+w_m).$$
We will need the following two limits, which both follow from $\linf (1-\phi_m)=0$:
        \begin{align}\linf \frac{w_{m}}{x_{m}(x_{m}+w_{m})}= \linf\frac{\sum_{i=2}^{\infty}\frac{(1-\phi_{c}(m))^{i-1}}{i}}{\sum_{i=1}^{\infty}m\frac{(1-\phi_{c}(m))^i}{i}}=\linf\frac{\sum_{i=2}^{\infty}\frac{(1-\phi_{c}(m))^{i-2}}{i}}{\sum_{i=1}^{\infty}m\frac{(1-\phi_{c}(m))^{i-1}}{i}}=0. \end{align}%
    \begin{align}\linf \frac{w_{m}}{x_{m}}= \linf\sum_{i=2}^{\infty}\frac{(1-\phi_{c}(m))^{i-1}}{i}=0.\end{align}%

From this we can conclude that  $\frac{w_{m}s}{x_{m}(w_{m}+x_{m})\times\sum_{i=1}^{\infty} \frac{(s(x_{m}+w_{m}))^{i-1}}{i!}}$ converges uniformly against $0$ (independent of $s$; recall that $s\in (0,1]$), as
 \begin{align*}
     0\leq \frac{w_{m}s}{x_{m}(w_{m}+x_{m})\times\sum_{i=1}^{\infty} \frac{(s(x_{m}+w_{m}))^{i-1}}{i!}}\leq  \frac{w_{m}s}{x_{m}(w_{m}+x_{m})}\leq \frac{w_{m}}{x_{m}(w_{m}+x_{m})}
 \end{align*}
  and trivially $\frac{w_{m}+x_{m}}{x_{m}}$ converges uniformly against $1$.   By \Cref{technical} $\gamma(s,x)$ is uniformly continuous over $(0,1]\times \mathbb{R}$. This means that for any $\epsilon>0$, there exists a $\delta>0$ such that if $|x-x'|<\delta$ then $|\gamma(s,x)-\gamma(s,x')|<\epsilon$ for all $s\in(0,1]$.
 So since $\linf x_{m}+w_{m}=L$, $\gamma(s,x_{m}+w_{m})$ converges uniformly against $\gamma(s,L)$. We conclude that $\frac{1}{x_{m}}-\frac{s}{e^{s(x_{m}+w_{m})}}$ converges uniformly (in terms of $s$) against $\gamma(s,L)$.
So for any $\epsilon>0$ and for any $s$ there is some $m_0$, such that for $m\geq m_0$
   \begin{align*}&\gamma(s,L)-\epsilon\leq \frac{1}{x_{m}}-\frac{s}{e^{s(x_{m}+w_{m})}}\leq \gamma(s,L) +\epsilon, \end{align*}
   and in particular for all $i\in [m]$
   \begin{align*}&\gamma(s(i),L)-\epsilon\leq \frac{1}{x_{m}}-\frac{s}{e^{s(x_{m}+w_{m})}}\leq \gamma(s(i),L) +\epsilon, \end{align*}
   and hence
   \begin{align*}&\frac{1}{m}\sum_{i=1}^{m}(\gamma(s(i),L)-\epsilon)\leq \frac{1}{m}\sum_{i=1}^{m} (\frac{1}{x_{m}}-\frac{s}{e^{s(x_{m}+w_{m})}})\leq \frac{1}{m}\sum_{i=1}^{m}(\gamma(s(i),L) +\epsilon)  \end{align*}
   implying the first equality in Equation \ref{integral}. 
By \Cref{technical}, $\gamma(s,x)$ is strictly decreasing in $x$ and hence $4\cdot\int_{0}^1 \gamma(s,x) ds$ is strictly decreasing in $x$ too, implying in particular that for each $c\in[0,1]$, there is a unique $x$ satisfying the equality.
We have shown that $\linf (1-\phi_m)\times m= L$ satisfies $4\cdot\int_{0}^1 \gamma(s,L) ds=c$, implying it is the unique solution, as desired,
\end{proof}

We remind the reader that $h^{\swap}$ is defined by $h^{\swap}(c)=\lim_{m\rightarrow \infty} (1-\phi_m^{\mathrm{swap}}(c))
\times m$.
\begin{lemma}\label{hswap}
    The function $h^{\swap}$ is a bijective strictly decreasing function from $(0,1]$ to $[0,\infty)$.
\end{lemma}
\begin{proof}
We  shown that $\lim_{m\rightarrow \infty} (1-\phi_m^{\mathrm{swap}}(c))
\times m=L$ is the unique solution to $\int_{0}^1 \gamma(s,L) ds=\frac{c}{2}$. We now evaluate this integral.
Using a change of variables and wolfram alpha, we obtain
\begin{align}&\int_{0}^1 \gamma(s,L) ds= \frac{1}{L}\int_{0}^1 1-sL\frac{1}{e^{sL}-1}ds\\
    &=\frac{1}{L}(1-\int_{0}^L z\frac{1}{e^z-1}\frac{dz}{L})\\
    &=\frac{1}{L}-\frac{1}{L^2}\int_{0}^L z\frac{1}{e^z-1}dz\\
    &=\frac{1}{L}-\frac{1}{L^2}[z \log(1-e^{-z})-Li_2(e^{-z})]]_0^L \label{wa} \\
    &=\frac{1}{L}-\frac{1}{L^2} (L \log(1-e^{-L})-Li_2(e^{-L})+Li_2(1))\\
    &=\frac{1}{L}- \frac{\log(1-e^{-L})}{L}+\frac{Li_2(e^{-L}))}{L^2}-\frac{Li_2(1)}{L^2},
\end{align}
where Equality \ref{wa} was verified using Wolfram Alpha.
where $Li_2(z)=\sum_{k=1}^{\infty}\frac{z^k}{k^2}$ is the Dilogarithm, a function strictly increasing on the reals.
Let $r(L)=4\int_{0}^1 \gamma(s,L)ds$ is strictly decreasing on $[0,\infty)$ with $r(0)=4\int_{0}^1 f(s,0)ds=4\int_{0}^1 \frac{1}{2}s ds=1$ and $\lim_{L\rightarrow \infty}=0$.
So $r$ maps $[0,\infty)$ bijectively to $[1,0)$ and is strictly decreasing. Since $h^{swap}(c)=L$, $h^{swap}$ is the inverse of $r$, $h^{swap}=r^{-1}$ and therefore strictly decreasing and maps $[1,0)$ bijectively to $[0,\infty)$, as required.
\end{proof}

\begin{proof}[Proof of Theorem \ref{thm:swapphi}]
Remember that for $c\in [0,1]$, $\phi_m^{\mathrm{swap}}(c)$ is defined as to satisfy
\begin{align}&g^{swap}_m(\phi_m^{\mathrm{swap}}(c))=\frac{4\cdot\mathbb{E}_{v\sim \calM_{\phi_m^{\mathrm{swap}}(c),m,v^*}}[\K(v,v^*)]}{m(m-1}=c\\
&\implies \lim_{m\rightarrow \infty}\frac{\mathbb{E}_{v\sim \calM_{\phi_m^{\mathrm{swap}}(c),m,v^*}}[\K(v,v^*)]}{m^2}=\frac{c}{2},\label{last}\text{ where }0<\frac{c}{2}<\frac{1}{4}.\end{align}
If $c=0$, $\phi_m^{\mathrm{swap}}=0$ for all $m$ and if $c=1$, then we must have $\phi_m^{\mathrm{swap}}=0$ for all $m$.

Now fix $1>c>0$. Let $x_{m}=m\times (1-\phi_m^{\mathrm{swap}}(c))$ and note that $x_m$ is non-negative for all $m$. 
Suppose $x_{k(m)}$ is a convergent subsequence of $x_m$ with limit $L\geq 0$.
By \Cref{lem:unique}, $L$ is the unique solution $x$ to $\int_{0}^1\gamma(s,x)ds=c$
and $\lim_{m\rightarrow \infty}g^{swap}_m(\phi_m^{\mathrm{swap}}(c))=\int_{0}^1\gamma(s,x)ds.$
By \Cref{technical}, $\int_{0}^1f(x,0)dx=\frac{1}{4}$,
so if $L=0$ then $$\lim_{m\rightarrow \infty}g^{swap}_m(\phi_m^{\mathrm{swap}}(c))=\linf \frac{4\cdot \mathbb{E}_{v\sim \calM_{\phi_m^{\mathrm{swap}}(c),m,v^*}}[\K(v,v^*)]}{m(m-1)}=1,$$ thereby contradicting
 $$\lim_{m\rightarrow \infty}g^{swap}_m(\phi_m^{\mathrm{swap}}(c))=c<1.$$ So $L>0$. We have shown that any convergent subsequence of $x_m$ must tend to $L>0$, the unique solution to $\int_{0}^1\gamma(s,x)ds=c$.
Suppose that $\linf x_m\neq L$. Then for there exists $\epsilon>0$ such that for all $m_0\in \mathbb{N}$, there exists $m\geq m_0$ with $|x_m-L|>\epsilon$ and in particular we obtain a subsequence $x_{k_m}$ of $x_m$ satisfying $|x_{k_m}-L|>\epsilon$ for all $m$.
Since $x_{m}$ was shown to be bounded, so is $x_{k_m}$, implying by Bolzano Weierstrass that it in turn has a convergent subsequence $x_{k'_m}$. This gives us a contradiction as $|x_{k'(m)}-L|>\epsilon$ for all $m$ and yet as it converges, so as a subsequence of $x_m$ it converges to $L$. We conclude that $\linf x_m = L$.
The statement about $h^{swap}$ is shown in \Cref{hswap}.

\end{proof}

\subsection{Missing Proofs from \Cref{sec:asy_cover}}
\symmetry*
\begin{proof}
Suppose that parameterizing by property $\mathcal{Y}$ asymptotically covers property $\mathcal{X}$. Then by definition $\linf g^{\mathcal{X}}_m(\phi^\mathcal{Y}_m(c))=f(c)$ for $c\in[0,1]$ where $f:[0,1]\mapsto [0,1]$ is strictly monotonic and bijective.
So $f^{-1}$ exists and we write $f^{-1}(d)=c$.
Since $g^{\mathcal{X}}_m$ and $\phi^\mathcal{Y}_m$ are continuous, and bijectively map $[0,1]$ to $[0,1]$, so  $(g^{\mathcal{X}}_m(\phi^Y_m)))^{-1}=g^{\mathcal{Y}}_m(\phi^\mathcal{X}_m)$ is continuous and bijecetively maps $[0,1]$ to $[0,1]$ too.
So by continuity, for $\epsilon>0$, there exists $\delta >0$ such that if $|x-x'|<\delta$ then $|g^{\mathcal{Y}}_m(\phi^\mathcal{X}_m)(\mathcal{X})-g^{\mathcal{Y}}_m(\phi^\mathcal{X}_m)(x')|<\epsilon$.
But for any $\delta >0$, there exists a large enough $m$ such that $|g^{\mathcal{X}}_m(\phi^\mathcal{Y}_m(c))-f(c)|<\delta$, so by continuity of $g^{\mathcal{X}}_m(\phi^\mathcal{Y}_m$,
\begin{align}&|g^{\mathcal{Y}}_m(\phi^\mathcal{X}_m(g^{\mathcal{X}}_m(\phi^\mathcal{Y}_m(c)))-g^{\mathcal{Y}}_m(\phi^\mathcal{X}_m((f(c)))|\\&=|c-g^{\mathcal{Y}}_m(\phi^\mathcal{X}_m)(f(c))|\\&=|f^{-1}(d)-g^{\mathcal{Y}}_m(\phi^\mathcal{X}_m)(d)|<\epsilon.\end{align}
We conclude that if parameterizing by property $\mathcal{Y}$ asymptotically covers $\mathcal{X}$, then also parameterizing by property $\mathcal{X}$ asymptotically covers $\mathcal{Y}$.

Now suppose that it is not the case, that parameterizing by property $\mathcal{Y}$ asymptotically cannot distinguish property $\mathcal{X}$.
Note that since $g^{\mathcal{X}}_m(\phi^\mathcal{Y}_m(c))$ is strictly monotonic and bounded, it converges to some $L\in[0,1]$.
Then by assumption it must hold that $\linf g^{\mathcal{X}}_m(\phi^\mathcal{Y}_m(c_1))=L_1$ $\linf g^{\mathcal{X}}_m(\phi^\mathcal{Y}_m(c_2))=L_2$ for some $ c_1,c_2,L_1,L_2\in[0,1]$, $c_1\neq c_2$ and $L_1\neq L_2$.
Then by continuity and bijectivity of $g^{\mathcal{X}}_m(\phi^\mathcal{Y}_m)$, we can see (e.g. from an analogous argument as above) that $\linf g^{\mathcal{Y}}_m(\phi^\mathcal{X}_m(L_1))=c_1$ and $\linf g^{\mathcal{Y}}_m(\phi^\mathcal{X}_m(L_2))=c_2$.
This implies that it is also does not hold that parameterizing by property $\mathcal{X}$ cannot asymptotically distinguish $\mathcal{Y}$.
\end{proof}
The following Proposition \ref{normphilim} is used in the proof of Theorem \ref{th:impli}.
\begin{proposition}\label{normphilim}
    Let $\epsilon\in (0,1]$.
    Then $\linf \phi_{m}^{\mathcal{\swap}}(\epsilon)=1$.
\end{proposition}
\begin{proof}
    Since by Fact \ref{strict}, $\mathbb{E}_{v\sim \calM_{\phi,m,v^*}}[\K(v,v^*)]$ is strictly increasing and continuous in $\phi$, for all $m\in \mathbb{N}$, $g_{m}^{\mathcal{\swap}}(\phi)$ is strictly increasing and continuous in $\phi$ and furthermore $g_{m}^{\mathcal{\swap}}(0)=0$ and $g_{m}^{\mathcal{\swap}}(1)$. Since strict monotonicity is preserved under taking the inverse of a function, $\phi_{m}^{\mathcal{\swap}}=(g_{m}^{\mathcal{\swap}})^{-1}$ is strictly increasing on $[0,1]$.
    As $\phi_{m}^{\mathcal{\swap}}(\epsilon)$ is bounded in $[0,1]$ for all $m\in \mathbb{N}$ and strictly increasing, it follows by the monotone convergence theorem that $\linf \phi_{m}^{\mathcal{\swap}}(\epsilon)= \phi\in[0,1]$ exists. We must have $\phi=1$, as for $\phi<1$ by Corollary  \ref{Cor:zero} the expected normalized swap distance is $0$, contradicting $0<\epsilon$.
\end{proof}

\implications*
\begin{proof}
\textbf{Proof of (1):} Suppose $g_{m}^{\mathcal{X}}$ is strictly increasing (so $f$ is too). By bijectivity of $f$, $f(0)=0$,
and so
for any $\phi\in[0,1)$, $$f(0)\leq \linf g_{m}^{\mathcal{X}}(\phi)\leq \linf g_{m}^{\mathcal{X}}(\phi_{m}^{\mathcal{\swap}}(\epsilon))=f(\epsilon)$$ where  $\epsilon\in (0,1]$. The inequality follows since  $\linf \phi_{m}^{\mathcal{\swap}}(\epsilon)=1$ by \Cref{normphilim} and because  $g_{m}^{\mathcal{X}}$ is strictly increasing.  We conclude that $\linf g_{m}^{\mathcal{X}}(\phi)=f(0)$ by continuity of $f$. The case that $g_{m}^{\mathcal{X}}$ is strictly decreasing is analogous.

\textbf{Proof of (2):} Suppose $g_{m}^{\mathcal{X}}$ is strictly increasing (so $f$ is too), then for any $c\in(0,1]$, $$f(1)\geq \linf g_{m}^{\mathcal{X}}(\phi_{m}^{\mathcal{\swap}}(c))\geq \linf g_{m}^{\mathcal{X}}(1-\epsilon)=f(1-\epsilon)$$ for any $\epsilon \in (0,1]$ again because $\linf \phi_{m}^{\mathcal{\swap}}(c)=1$ by \Cref{normphilim}. We conclude that $\linf g_{m}^{\mathcal{X}}(\phi_{m}^{\mathcal{\swap}}(c))=f(1)$ by continuity of $f$. The case that $g_{m}^{\mathcal{X}}$ is strictly decreasing is analogous.

\end{proof}
\subsection{Missing Proofs from \Cref{sec:c1}}
\exppos*
\begin{proof}
    Let $\phi\in [0,1)$. Since Fact \ref{le:c1} says that the position of alternative $c_1$ is distributed according to a truncated geometric distribution with parameters $m$ and $(1-\phi)$, we have that the expected value of alternative $c_1$'s position is
    \begin{align}&\mathbb{E}_{v\sim \calM_{\phi,m}}[\pos(v,c_1)]=\sum_{i=1}^{m}i \cdot\frac{\phi^{i-1}}{\sum_{i=1}^m \phi^{i-1}}
    =\frac{\sum_{i=1}^m i\cdot \phi^{i-1}}{\sum_{i=1}^m \phi^{i-1}}\label{1}\\&=\frac{\frac{1-\phi^{m+1}}{(1-\phi)^2}-(m+1)\frac{\phi^m}{1-\phi}}{\frac{1-\phi^{m}}{1-\phi}}=\frac{1}{1-\phi}-m\frac{\phi^m}{1-\phi^m}\label{2}\end{align}
    where Line \ref{2} uses the geometric sum and its derivative.
\end{proof}

\expposnorm*
\begin{proof}
From \Cref{hswap} and \Cref{loglimit}, we know that for fixed $c\in(0,1]$: 
\begin{align}
    &\linf (1-\phi_m^{\mathrm{swap}}(c))\times m=h^{\swap}(c)>0\label{ri1a}\\
     &\linf \phi_m^{\mathrm{swap}}(c)^ m=e^{-h^{\swap}(c)}<1\label{ri2a}
\end{align}

Using these we can conclude that:
\begin{align*}
    & \linf g_m^{\poso}(\phi_{m}^{\swap}(c))\\=
    & \linf \frac{\frac{1}{1-\phi_{m}^{\swap}(c)}-m\frac{\phi_{m}^{\swap}(c)^m}{1-\phi_{m}^{\swap}(c)^m}-1}{\frac{m-1}{2}+1}\\=
    & \linf 2\cdot (\frac{1}{m(1-\phi_{m}^{\swap}(c))}-\frac{\phi_{m}^{\swap}(c)^m}{1-\phi_{m}^{\swap}(c)^m})\\=
    &  2 \cdot (\frac{1}{\linf m(1-\phi_{m}^{\swap}(c))}-\frac{1}{-1+\linf \phi_{m}^{\swap}(c)^{-m}})\\=
    & 2 \cdot (\frac{1}{h^{\swap}(c)}-\frac{1}{e^{h^{\swap}(c)}-1}):=f(c).
\end{align*}
\paragraph{Properties of $f(c)$}
It follows from \Cref{technical} that $t(x)=\frac{1}{x}-\frac{1}{e^x-1}$ is strictly decreasing and that $t(0)=\frac{1}{2}$. Furthermore, $\lim_{x\rightarrow \infty}t(x)=0$, so that $t$ maps  $[0,\infty)$ bijectively to $[\frac{1}{2},0)$.
By Lemma \ref{hswap}, $h^{swap}$ is strictly decreasing and bijectively maps $[0,1)$ to $[0,\infty)$.
Since compositions of strictly decreasing functions are strictly decreasing $f(c)=2\cdot t(h^{swap})(c)$ is strictly decreasing and furthermore $f(c)$ maps $[0,1)$ bijectively to $[0,1)$.

\end{proof}

\subsection{Missing Proofs from \Cref{sec:PC}}

\comparingTwo*
\begin{proof}
Denote by $q_{i,j}$ the probability that alternative $c_i$ is ranked before alternative $c_j$ in $v$ and denote by $p^m_{i,j}$ the probability that alternative $c_i$ is ranked in position $j$ in $v$ when $v\sim \mathcal{M}_{\phi,m}$.
\citet{mallows1957non} showed for fixed $\phi\in[0,1]$ that $q_{i,j}$ is independent of $m$ and only depends on the relative difference $j-i$. So for $k=j-i+1$, it holds that $q_{i,j}=q_{1,k}$.
Using the Random Insertion sampling procedure for Mallows as discussed in \Cref{sec:Deleting}, we can calculate $q_{1,k}$. We only need to consider the iteration during which alternative $c_k$ is inserted and reason about which position alternative $c_1$ is in when this happens and the probability that $c_k$ is inserted below alternative $c_1$. Before $c_k$ is inserted, alternative $c_1$ is ranked in position $i$ with probability $\frac{\phi^{i-1}}{\sum_{j=0}^{k-1}\phi^j}$ by Fact \ref{le:c1}. 
The probability that $c_k$ is inserted in position $j$ is $\phi^{k-j}$. If $c_1$ is ranked in position $i$, then $c_k$ is inserted below position $i$ with probability $\frac{\sum_{j=0}^{k-i-1} \phi^j}{\sum_{j=0}^{k-1}\phi^j}$. So

$$q_{1,k}=\sum_{i=1}^{k-1}p^{k-1}_{1,i}\frac{\sum_{j=0}^{k-i-1} \phi^j}{\sum_{j=0}^{k-1}\phi^j}$$
Evaluating this we obtain the desired closed form expression:
        \begin{align*}&q_{1,k}=\sum_{i=1}^{k-1}p^{k-1}_{1,i}\frac{\sum_{j=0}^{k-i-1} \phi^j}{\sum_{j=0}^{k-1}\phi^j}=\sum_{i=1}^{k-1} \frac{(1-\phi)\phi^{i-1}}{1-\phi^{k-1}}\frac{1-\phi^{k-i}}{1-\phi^k}
         \\&=\frac{(1-\phi)}{(1-\phi^{k})(1-\phi^{k-1})}\sum_{i=1}^{k-1}\phi^{i-1}(1-\phi^{k-i})
         \\&=\frac{(1-\phi)}{(1-\phi^{k})(1-\phi^{k-1})}(\sum_{i=1}^{k-1}\phi^{i-1} - \sum_{i=1}^{k-1}\phi^{k-1})
         \\ &=\frac{(1-\phi)}{(1-\phi^{k})(1-\phi^{k-1})}(\frac{1-\phi^{k-1}}{1-\phi} - (k-1)\phi^{k-1})=\frac{1}{1-\phi^k}(1-\frac{(1-\phi)(k-1)\phi^{k-1}}{1-\phi^{k-1}}).\end{align*}

\end{proof}
\promm*
\begin{proof}

From \Cref{hswap} and \Cref{loglimit}, we know that for fixed $c\in(0,1]$: 
\begin{align}
    &\linf (1-\phi_m^{\mathrm{swap}}(c))\times m=h^{\swap}(c)>0\label{ri1}\\
     &\linf \phi_m^{\mathrm{swap}}(c)^ m=e^{-h^{\swap}(c)}<1\label{ri2}
\end{align}

Using these we can conclude that:
\begin{align*}&\linf g_m^{( 1\text{ beats }m)}(\phi_{m}^{\swap}(c))\\&=
\linf  \frac{1}{1-\phi_{m}^{\swap}(c)^m}-\frac{(1-\phi_{m}^{\swap}(c))(m-1)\phi_{m}^{\swap}(c)^{m-1}}{(1-\phi_{m}^{\swap}(c)^m)(1-\phi_{m}^{\swap}(c)^{m-1})}\\&=
\frac{1}{1-\linf \phi_{m}^{\swap}(c)^m}-\frac{\linf (1-\phi_{m}^{\swap}(c))(m-1)\cdot \linf \phi_{m}^{\swap}(c)^{m-1}}{\linf (1-\phi_{m}^{\swap}(c)^m) \cdot \linf (1-\phi_{m}^{\swap}(c)^{m-1})}\\&=\frac{1}{1-e^{h^{\swap}(c)}}-\frac{h^{\swap}(c)e^{h^{\swap}(c)}}{(1-e^{-h^{\swap}(c)})^2}=
\frac{1}{1-e^{-h^{\swap}(c)}}(1-\frac{h^{\swap}(c)}{e^{h^{\swap}(c)}-1}):=f(c),\end{align*}
where the first equality holds because of \Cref{winprob}, the second because of the algebraic limit theorem, and the third because of \Cref{ri1,ri2}.

\paragraph{Properties of $f$.}
    We show that the function $t(x)=2\cdot \frac{1}{1-e^{-x}}(1-\frac{x}{e^x-1})-1$ is strictly increasing and maps $[0,\infty)$ bijectively to $[\frac{1}{2},1)$. The derivative of $t$ is $\frac{d}{dx}t(x)= \frac{e^x (e^x (x-2) + x+2)}{(e^x-1)^3}$ which is positive for $x>0$ since $(e^x (x-2) + x+2)$ is strictly increasing and equals $0$ if $x=0$. So $t$ is strictly increasing on $(0,\infty)$, with $\lim_{x\rightarrow \infty} t(x)=2 \cdot (1-0)-1=1$.
    To evaluate  $t$ at $x=0$, note that $t$ can be rewritten as $t(x)=2\cdot\frac{e^x(e^x-1-x)}{(e^x-1)^2}-1$.
    Then using the Taylor expansion of $e^x$ \begin{align*}
    &t(x)=2\cdot \frac{e^x(\sum_{i=0}\frac{x^i}{i!}-x-1)}{(\sum_{i=0}\frac{x^i}{i!}-1)^2}-1\\&=2\cdot \frac{e^x(\sum_{i=2}\frac{x^i}{i!})}{(\sum_{i=1}\frac{x^i}{i!})^2}-1=2\cdot \frac{e^x(\sum_{i=2}\frac{x^{i-2}}{i!})}{(\sum_{i=1}\frac{x^{i-1}}{i!})^2}-1\\&=2\cdot \frac{1\cdot\frac{1}{2}}{1^2}-1=0\text{ if $x=0$.}\end{align*}

    Since $h^{swap}$ is strictly decreasing and maps $(0,1]$ bijectively to $[0,\infty)$ by Lemma \ref{hswap} and $t$ is strictly increasing, mapping $[0,\infty)$ bijectively to $[0,1)$, we conclude that $f(c)=t(h^{swap(c)})$ is strictly decreasing, mapping $(0,1]$ bijectively to $[0,1)$
\end{proof}

\begin{figure*}[t]
\centering
\begin{subfigure}{0.35\textwidth}
  \centering
  \resizebox{0.9\textwidth}{!}{\input{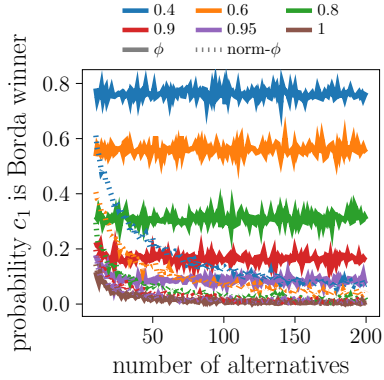}}
  \caption{Probability that $c_1$ is Borda winner.}\label{fig:winne1}
\end{subfigure}\quad\quad
\begin{subfigure}{0.35\textwidth}
  \centering
  \resizebox{0.9\textwidth}{!}{\input{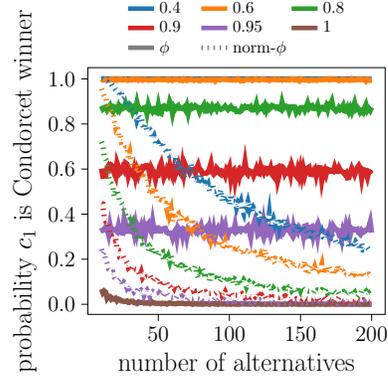}}
  \caption{Probability that $c_1$ is Condorcet winner.}\label{fig:winne2}
\end{subfigure}
\caption{Influence of the number of alternatives $m$ on the probability that $c_1$ is a Condorcet/Borda winner in ranking profiles sampled from the Mallows model for fixed values of the classical dispersion parameter $\phi$ (solid) and the normalized dispersion parameter $\normphi$ (dashed). Recall that $\phi=\normphi=0$ and $\phi=\normphi=1$, so the respective lines overlap. For each value of $m$, we sampled $1000$ profiles and computed the average results.}\label{fig:winners}
\end{figure*}

\subsection{Additional Material for \Cref{sec:Winners}}\label{app:winners}
In the main body in \Cref{fig:elecsub3}, we have seen how the probability that $c_1$ is the Plurality winner develops when varying the number of alternatives. 
We have observed that for a fixed value of $\phi$ in the classical Mallows model this probability remains roughly constant, whereas for a fixed value of $\normphi$ in the normalized Mallows model it decreases when increasing $m$. 
In this section, we look at two further voting rules: Borda and Condorcet. 
Under the Borda voting rule, each ranking awards $m-i$ points to the alternative it ranks in $i$th place and the alternative with the highest number of points wins. 
Further, an alternative $c$ is a Condorcet winner if for each alternative $c'\neq c$ a majority of rankings rank $c$ before $c'$.
We depict the dependency of the probability that $c_1$ is a Condorcet/Borda winner on the number of alternatives for fixed values of $\phi$/$\normphi$ in \Cref{fig:winners}.
It turns out that the situation for Borda and Condorcet is analogous to the picture for the Plurality voting rule:
For a fixed value of $\phi$ the probability that $c_1$ is a winner is kept constant, whereas for a fixed value of $\normphi$ it decreases when increasing the number of alternatives.

\newpage

\begin{figure}[t]
\centering
\begin{subfigure}[b]{.35\textwidth}
  \centering
  \resizebox{.9\textwidth}{!}{\input{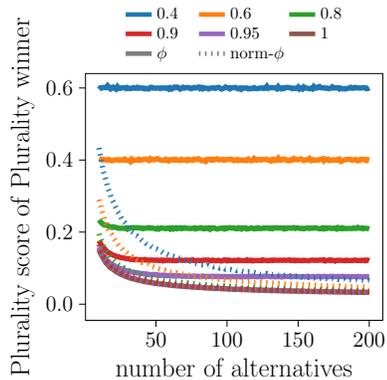}}
  \caption{Normalized plurality score of Plurality winner.}
\end{subfigure}\quad\quad
\begin{subfigure}[b]{.35\textwidth}
  \centering
  \resizebox{.9\textwidth}{!}{\input{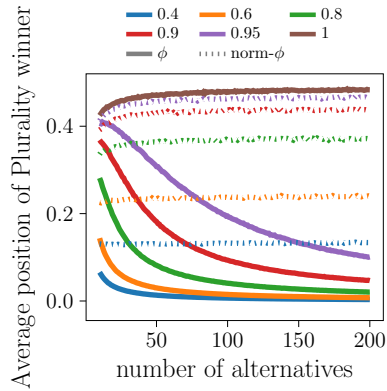}}
  \caption{Average Position of Plurality Winner.\\ \phantom{.}}
\end{subfigure}
\vspace{4mm}

\begin{subfigure}[b]{.35\textwidth}
  \centering
  \resizebox{0.9\textwidth}{!}{\input{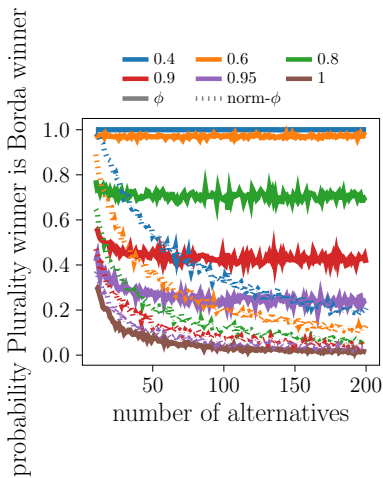}}
  \caption{Probability that Plurality Winner is Borda Winner.}
\end{subfigure}\quad\quad
\begin{subfigure}[b]{.35\textwidth}
  \centering
  \resizebox{0.9\textwidth}{!}{\input{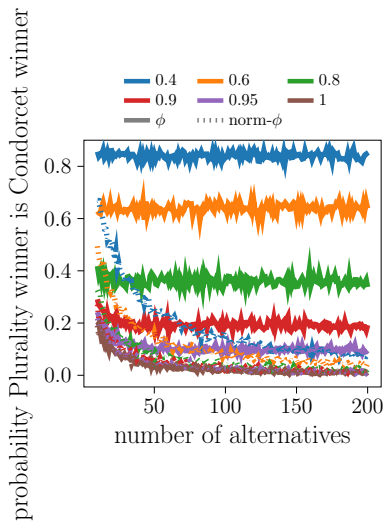}}
  \caption{Probability that Plurality Winner is Condorcet winner.}
\end{subfigure}
\caption{Properties of profiles with $n=100$ rankings and a varying number of alternatives sampled from Mallows model with classical dispersion parameter $\phi$ (top) or  normalized dispersion parameter $\normphi$ (bottom). For each value of $m$, we sampled $1000$ profiles and computed the average results.}
\label{fig:profilesproperties}
\end{figure}

\begin{table*}[t]
    \centering
    \resizebox{\textwidth}{!}{\begin{tabular}{l|c|c|c}
         & FBS systems & restricted FCS\&FBS systems  & FCS\&FBS systems \\ \hline
        Plurality score of Plurality winner ($\searrow_{\text{norm}}$) & 0.59 & 0.63&  0.49 \\
        Average position of Plurality winner  ($\rightarrow_{\text{norm}}$)& 0.012 & 0.01 & 0.011 \\
        Fraction of profiles where Borda and Plurality winner coincide ($\searrow_{\text{norm}}$) & 0.85 & 0.85& 0.6 \\
        Fraction of profiles where Plurality and Condorcet winner coincide ($\searrow_{\text{norm}}$) & 0.91 & 0.91 & 0.72
    \end{tabular}}
    \caption{Average values of properties in American football profiles.}
    \label{tab:football}
\end{table*}

\section{Additional Material for \Cref{sub:evidence}}\label{app:real-worldData}
In \Cref{sub:posDisID}, we have observed that in real-world profiles with varying numbers of alternatives, the positionwise distance to ID stays roughly constant as in the normalized Mallows model for a fixed value of $\normphi$.
In this section, we extend this analysis to some of the properties we have considered in \Cref{sec:fixvsnorm}. 
All of these properties deal with specific candidates from the central order, e.g., candidate $c_1$ who appears in the first position of the central order and candidate $c_m$ who appears in the last position. 
In real-world profiles, there is naturally no central order in the sense of the central order of the Mallows model.
To still be able to compare the behavior of profiles sampled from the Mallows model and real-life profiles, we thus focus on properties only involving $c_1$ and use the Plurality winner as a proxy for $c_1$.
This is a natural approach, as $c_1$ is the candidate which has the highest probability to be ranked first in a vote sampled from the Mallows model. 
To analyze the implications of replacing $c_1$ by the Plurality winner of some sampled profile, we rerun some of our experiments. 
In \Cref{fig:profilesproperties}, we present the behavior of the (normalized) Mallows model with a fixed (normalized) dispersion parameter concerning properties involving the Plurality winner when varying the number of alternatives.
We sample profiles containing $100$ rankings and analyze different properties of the Plurality winner analogous to our previous analysis of alternative $c_1$.
It turns out that examining properties of the Plurality winner instead of alternative $c_1$ does not lead to a significant change in the results (see the similarities to \Cref{fig:elecsub1,fig:elecsub2,fig:winne1,fig:winne2}). 
In particular, as before the classical Mallows model keeps the fraction of rankings in which the Plurality winner is ranked first roughly constant as well as the probability that the Plurality winner is the Borda/Condorcet winner.
In contrast, the normalized Mallows model keeps the average position of the Plurality winner constant. 

We now turn to analyzing how these properties behave in real-world profiles. 
As in contrast to the positionwise distance from ID, some of these properties are binary and for all of them we observe a high fragility, we want to average over the behavior of different profiles to get a clear picture. 
As for the Tour de France and Spotify profiles, there is no natural grouping of profiles that allow for some natural averaging, we do not look at them in this part. 
Instead, we focus on the American Football profiles for which there is a natural split into two (large) groups containing profiles over roughly the same number of alternatives: The profiles for outlets that rank only the FBS teams (in those profiles the number of candidates lies between $118$ and $132$) and profiles for outlets ranking FCS\&FBS teams (in those profiles the number of candidates lies between $239$ and $258$). Averaging over the results within the two groups allows for a more robust estimate of the quantities. 
We depict the results in \Cref{tab:football}. 
As a sanity check we also added quantities for the FCS\&FBS profiles restricted to the FBS teams (in order to check whether the observed trends are due to the changed number of alternatives or due to a fundamental difference how the two groups of outlets generate their rankings). 

For all properties that are kept constant by the classical Mallows model, we see a significant decrease when moving from the American Football profiles over around $~125$ to the profiles over around $~245$ alternatives. 
In contrast, the average normalized position of the Plurality winner, which is kept constant by the normalized Mallows model, also stays constant when doubling the number of alternatives in the real-world profiles. 
These experiments underline that the behavior of the normalized Mallows model for a fixed normalized dispersion parameter is much more in line with what is present in real-world profiles, wheras the classical Mallow model behaves differently.

\newpage

\section{Additional Material for \Cref{sec:Deleting}}\label{app:Deleting}
\begin{figure}[h!]
\centering
\begin{subfigure}[b]{.35\textwidth}
  \centering
  \resizebox{.9\textwidth}{!}{\input{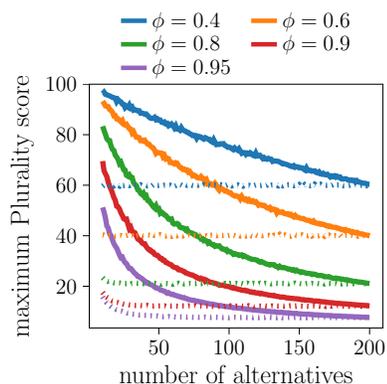}}
  \caption{Average Plurality Score of Plurality\\ Winner.}
\end{subfigure}\quad\quad
\begin{subfigure}[b]{.35\textwidth}
  \centering
  \resizebox{.9\textwidth}{!}{\input{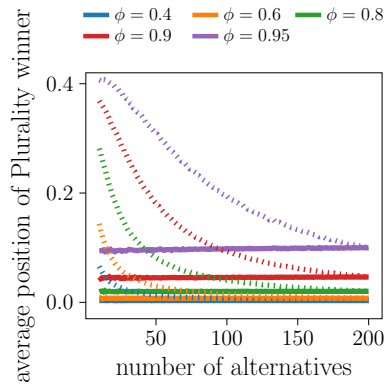}}
  \caption{Average Position of Plurality Winner.\\ \phantom{.}}
\end{subfigure}
\vspace{4mm}

\begin{subfigure}[b]{.35\textwidth}
  \centering
  \resizebox{.9\textwidth}{!}{\input{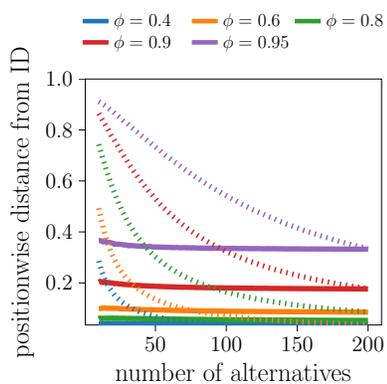}}
  \vspace{8.25mm}
  \caption{Positionwise Distance from ID.\\ \phantom{.}}
\end{subfigure}\quad\quad
\begin{subfigure}[b]{.35\textwidth}
  \centering
  \resizebox{.9\textwidth}{!}{\input{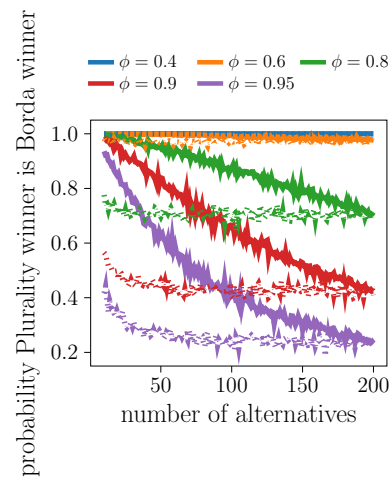}}
  \caption{Probability that Plurality Winner is Borda winner.}
\end{subfigure}

\caption{
Analysis of how the properties of profiles containing $100$ rankings depend on the number of alternatives for two different ways of sampling from the classical Mallows model. 
We compare sampling profiles with a varying number of $m$ (dashed) with sampling profiles for $m=200$ alternatives and subsequently deleting some alternatives uniformly at random (solid).
}
\label{fig:delphi}
\end{figure}

\begin{figure}[h!]
\centering
\begin{subfigure}[b]{.35\textwidth}
  \centering
  \resizebox{.9\textwidth}{!}{\input{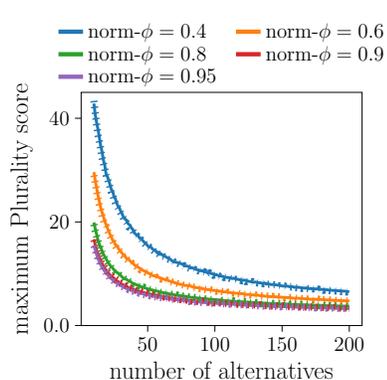}}
  \caption{Average Plurality Score of Plurality\\ Winner.}
\end{subfigure}\quad\quad
\begin{subfigure}[b]{.35\textwidth}
  \centering
  \resizebox{.9\textwidth}{!}{\input{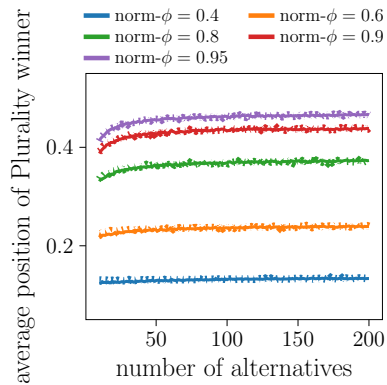}}
  \caption{Average Position of Plurality Winner.\\ \phantom{.}}
\end{subfigure}
\vspace{4mm}

\begin{subfigure}[b]{.35\textwidth}
  \centering
  \resizebox{.9\textwidth}{!}{\input{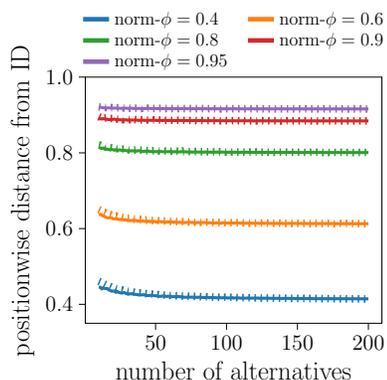}}
  \vspace{12mm}
  \caption{Positionwise Distance from ID.\\ \phantom{.}}
\end{subfigure}\quad\quad
\begin{subfigure}[b]{.35\textwidth}
  \centering
  \resizebox{.9\textwidth}{!}{\input{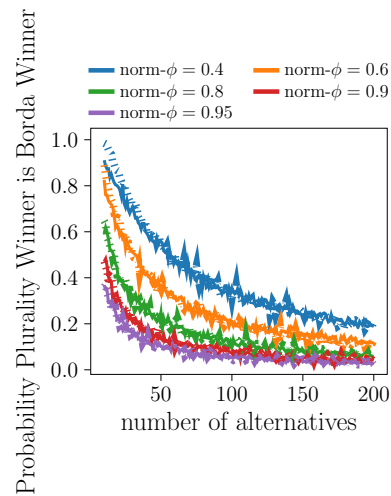}}
  \caption{Probability that Plurality Winner is Borda winner.}
\end{subfigure}

\caption{
Analysis of how the properties of profiles containing $100$ rankings depend on the number of alternatives for two different ways of sampling from the normalized Mallows model. 
We compare sampling profiles with a varying number of $m$ (dashed) with sampling profiles for $m=200$ alternatives and subsequently deleting some alternatives uniformly at random (solid).
}
\label{fig:delnormphi}
\end{figure}

In \Cref{sec:Deleting}, we have discussed the difference between sampling a profile for some number of $m$ of alternatives directly for some value of $\phi/\normphi$ or sampling a profile for some larger number of alternatives $i+m$ for the same value of $\phi/\normphi$ and then deleting $i$ alternatives uniformly at random. 
We have argued that for the normalized Mallows model (keeping $\normphi$ fixed) these two strategies result in profiles with very similar properties, whereas for the classical Mallows model (keeping $\phi$ fixed) the properties of the resulting profiles substantially differ. 
To support this claim we rerun the experiment described in \Cref{sec:Deleting}, for other properties of the sampled profile. 
Specifically, for the average position of the Plurality winner, the positionwise distance from ID, and the probability that the Plurality winner is the Borda winner (note that as in \Cref{app:real-worldData} we focus on the role of the Plurality winner instead of $c_1$ as the profiles sampled via the deletion strategy are formally not sampled from some Mallows distribution).
The results for the classical Mallows model can be found in \Cref{fig:delphi} and the results for the normalized Mallows model in \Cref{fig:delnormphi}. 
These additional results are in line with what we have observed in \Cref{sec:Deleting} and  confirm our intuition that for the normalized Mallows model deleting alternatives uniformly at random leads to ranking similar to those sampled for a smaller number of alternatives with the same value of $\normphi$, whereas this is clearly not the case for the classical Mallows model.

\end{document}